\documentclass[journal]{IEEEtran}
\usepackage{amsmath,amssymb}
\usepackage{amsfonts}
\usepackage{amsbsy}
\usepackage{graphicx}

\usepackage{subfigure}
\usepackage{epstopdf}
\usepackage{lipsum}
\usepackage{multirow}
\usepackage{dsfont}
\usepackage{xcolor}
\usepackage{blindtext}
\usepackage{balance}

\begin{document}

\title{Wireless Information and Power Transfer for IoT Applications in Overlay Cognitive Radio Networks}
\author{{Devendra~S.~Gurjar,~\IEEEmembership{Member,~IEEE}, Ha~H.~Nguyen,~\IEEEmembership{Senior Member,~IEEE}, and Hoang D. Tuan}
\thanks{D. S. Gurjar and H. H. Nguyen are with the Department of Electrical and Computer Engineering, University of Saskatchewan,
Saskatoon, SK S7N 5A9, Canada (e-mails: \{devendra.gurjar, ha.nguyen\}@usask.ca).}
\thanks{H. D. Tuan is with the Faculty of Engineering and Information Technology, University of Technology Sydney, Ultimo, NSW 2007,
Australia (e-mail: tuan.hoang@uts.edu.au).}
\thanks{This work was supported in part by an NSERC Discovery Grant 249772-2012.}}

%

\maketitle

\begin{abstract}
This paper proposes and investigates an overlay spectrum sharing system in conjunction with the simultaneous wireless information and power transfer (SWIPT) to enable communications for the Internet of Things (IoT) applications. Considered is a cooperative cognitive radio network, where two IoT devices (IoDs) exchange their information and also provide relay assistance to a pair of primary users (PUs). Different from most existing works, in this paper, both IoDs can harvest energy from the radio-frequency (RF) signals received from the PUs. By utilizing the harvested energy, they provide relay cooperation to PUs and realize their own communications. For harvesting energy, a time-switching (TS) based approach is adopted at both IoDs. With the proposed scheme, one round of bidirectional information exchange for both primary and IoT systems is performed in four phases, i.e., one energy harvesting (EH) phase and three information processing (IP) phases. Both IoDs rely on the decode-and-forward operation to facilitate relaying, whereas the PUs employ selection combining (SC) technique. For investigating the performance of the considered network, this paper first provides exact  expressions of user outage probability (OP) for the primary and IoT systems under Nakagami-$m$ fading. Then, by utilizing the expressions of user OP, the system throughput and energy efficiency are quantified together with the average end-to-end transmission time. Numerical and simulation results are provided to give useful insights into the system behavior and to highlight the impact of various system/channel parameters.
\end{abstract}


\section{Introduction}
\IEEEPARstart{S}{pectrum} sharing for the Internet of Things (IoT) is one of the most promising technologies in the fifth-generation (5G) wireless networks, which allows autonomous devices to realize communications for IoT applications in the licensed spectrum \cite{Khan2017}, \cite{Ercan2018}. The concept of IoT has been introduced with a vision to accommodate various physical things such as sensors, mobile phones, home appliances, healthcare gadgets, and even intelligent furniture, that can be connected through a communication network to exchange information about themselves and their surroundings. From the communications point of view, all these things, connected through a network, can be referred to as IoT devices (IoDs). Addressing communication aspects of such autonomous things (electronic devices) is crucial for bonding them together to form the IoT. Many applications are emerging to exploit the features and capabilities of IoT. For example, instruments can collaborate with each other in factories and farms to enhance the performance and efficiency of factory and farm operations \cite{Liu2017}. Exploiting IoT can also be useful in refineries where devices and sensors can be deployed to form automation in various operations without changing the core environment. Likewise, smart homes are made possible by implementing IoT-based home appliances. Moreover, IoDs are expected to be the critical entity for improving traffic management and transportation safety in autonomous driving vehicle industry \cite{Gharbieh2017}. 
For a reliable and ubiquitous IoT deployment, two fundamental challenges, i.e., network lifetime and spectrum scarcity, need to be addressed and they are the focus of this paper.

To prolong the lifetime of wireless communication networks, energy harvesting (EH) from the surrounding environment has been envisioned as one of promising solutions to counterpoise power limitations of connected wireless devices. Specifically, it has been observed that the conventional sources for EH such as solar, wind, thermoelectric, etc., could be unreliable to provide perpetual energy supply as these methods rely on location specific climate and environment \cite{Varshney}, \cite{Nasir}. Consequently, simultaneous wireless information and power transfer (SWIPT) technology is gaining tremendous attention due to its ability in providing sustainable and ubiquitous communications for numerous wireless communication scenarios, including IoT.
This technique exploits the idea that the radio-frequency (RF) signals can be utilized for both wireless power transfer and wireless information transfer at the same time \cite{Nasir}, \cite{Zhou}. Specifically, the antenna of a receiving node first captures the transmitted energy in RF radiation. Then, using an appropriate circuit \cite{Zhou}, the captured energy can be stored in the battery of that node after converting it into the direct current (DC) form. For enabling SWIPT in wireless networks, three practical receiver designs have been investigated in the literature, namely, time switching (TS), power splitting (PS), and antenna switching (AS) \cite{LWang2017}. In TS-based SWIPT, a receiving node switches in time between information processing (IP) and EH.  In PS-based SWIPT, the node splits the power of the received signal for IP and EH. The AS is another way to enable SWIPT in a multi-antenna based system, whereby the strongest antennas are exploited for IP, and others are used for EH or vice-versa \cite{Benkhelifa2017}. Although the amount of harvested energy from RF signals is currently less as compared to other conventional sources such as solar energy, its ubiquitous availability (indoor, outdoor, day or night) makes it more promising for future IoT networks.


Spectrum scarcity is  another critical design constraint in massive IoT deployments. Enabling IoT communications in the industrial, scientific, and medical (ISM) band is not a long-lasting and feasible solution as most of the wireless technologies operating in this band, e.g., ZigBee, Wi-Fi, and Bluetooth cannot provide seamless communications with the desired quality of service (QoS) \cite{Liu2017}.
On the other hand, it's not feasible to rely on the licensed band communications due to the scarcity of spectrum and the presence of a massive number of devices in IoT. Therefore, a promising solution is to exploit communications over the licensed spectrum without degrading the performance of legitimate users. Cooperative spectrum sharing is the suitable mechanism for achieving such an attribute in the IoT networks. For enabling spectrum sharing in such systems, three strategies are commonly adopted in the literature, i.e., \emph{interweave}, \emph{underlay}, and \emph{overlay} \cite{Goldsmith2009}.
As such, the interweave spectrum sharing suffers from traffic pattern errors of the primary system, whereas underlay spectrum sharing must comply with strict interference threshold constraint based on instantaneous channel state information which may be difficult to acquire in practice. As a result, the overlay scheme adopted in this paper is more appealing for such IoT systems. With this scheme, the IoDs could provide an incentive to the PUs for spectrum access, i.e., PUs could get benefits of improved performance due to relay assistance, while in return, relaying IoDs can explore their own transmission opportunities.

To summarize, incorporating the SWIPT technology with cooperative cognitive radio networks can effectively solve technical problems related to lifetime and spectrum scarcity in massive IoT deployment.

\subsection{Prior Arts}
In recent years, SWIPT technique has attracted significant attention for its inclusion in the relay-based wireless systems \cite{TLi}-\cite{Do}. Specifically, the authors in \cite{TLi} and \cite{HLee} have considered one-way relay networks, where a relay node can harvest energy from the RF signals received from the source node. To improve the spectral efficiency, the authors in \cite{Men}-\cite{Hu} have utilized the concept of SWIPT with amplify-and-forward (AF) relaying strategy for two-way relay systems. In particular, the work in \cite{Men} has jointly optimized the problem of relay selection and power allocation for a SWIPT-enabled asymmetric two-way relay system. The authors in \cite{Du} have considered a TS-based SWIPT scheme at the relay node and investigated the outage performance, whereas, the authors in \cite{Hu} have solved the optimization problem concerning power splitting factor and relay processing matrix for such spectral efficient systems. Different from AF-based two-way systems, the authors in \cite{Peng} and \cite{Do} have focused on decode-and-forward (DF) relaying scheme for SWIPT-enabled bidirectional relay systems.

Besides, research works in \cite{Yin2014}-\cite{Yan2018} have incorporated the concept of SWIPT in the spectrum sharing based systems and cellular networks. Specifically, the authors in \cite{Yin2014} have proposed a cognitive radio network, where the secondary node can extract energy from RF signals and utilize it for transmitting its own message or providing relay cooperation in different time slots. In particular, the authors have provided optimal conditions to maximize the system throughput for two scenarios, namely, cooperative mode and non-cooperative mode. Different from \cite{Yin2014}, the secondary node can transmit both primary and secondary signals simultaneously in \cite{Wang2016} with the overlay mode. For this set up, the authors have derived exact expressions of outage probability (OP) for both primary and secondary systems over Rayleigh fading channels. Further, research works \cite{Im2015} and \cite{Yang2016} have adopted an underlay cognitive radio scenario with EH and analyzed the systems' outage performance. By extending the system model of \cite{Im2015} and \cite{Yang2016}, the authors in \cite{Kalamkar2017} have considered multiple primary transmitters and receivers and evaluated the outage and ergodic capacity performance for the secondary system in the presence of multiple primary interferences. Furthermore, the authors in \cite{Verma2017} have introduced one-way cooperative cognitive radio network (CCRN) with energy assisted DF relaying and investigated the OP and throughput for both systems. With the similar system model as in \cite{Verma2017}, the authors in \cite{Yan2017} have studied opportunistic relaying by employing a dynamic SWIPT protocol. The authors in \cite{Nguyen2018} have derived the expressions of OP for EH-enabled CCRN under Nakagami-$m$ fading. In contrast to the one-way CCRN \cite{Yin2014}-\cite{Nguyen2018},  a cognitive two-way relay network with EH has been investigated in \cite{Wang2014} under Rayleigh fading.

Recently, a few works \cite{Yang2018}-\cite{Yan2018} have exploited the benefits of EH for IoT applications in the licensed band. In particular,  the authors in \cite{Yang2018} have studied resource allocation for a machine-to-machine enabled cellular network with EH by focusing on two different strategies, i.e., nonorthogonal multiple access and time division multiple access. Further, in \cite{Huang2018}, software-defined networking has been proposed to optimize network management and to control EH for IoT applications. Very recently, the authors in \cite{Yan2018} have developed a mathematical framework for the design and analysis of relay-assisted underlay cognitive radio networks with EH and investigated the systems' outage performance.

Most of the works as discussed above either considered EH with underlay cognitive radio networks or one-way CCRN, where the information exchange for both systems is carried out unidirectionally. To the authors' best knowledge, no work has yet considered the concept of SWIPT in a cognitive radio network with bidirectional primary and IoT transmissions.

\subsection{Main Contributions}
Focusing on the critical constraints of IoT deployments, this paper proposes a DF-based two-way CCRN with EH to leverage the benefits of both spectrum sharing and SWIPT technique. Herein, a pair of IoDs harvests energy from the RF signals by applying the TS technique and utilizes the accumulated energy for providing relay cooperation and its own information exchange. Moreover, the proposed scheme improves the overall spectrum efficiency and resolves the crucial reliability issue for primary links by enabling relay assistance from two IoDs in consecutive phases. For relaying, a DF-based operation is considered at the IoDs. With DF relaying, a selection combining (SC) technique is employed at the PUs to exploit multiple copies of their intended signals. The major contributions of this paper are summarized as follows:
\begin{itemize}
	\item This paper introduces an overlay spectrum sharing scheme with EH to enable IoT communications in the licensed spectrum.
	\item With the proposed scheme, exact expressions of user OP for both primary and IoT systems are derived under Nakagami-$m$ fading. Then, the expression of system throughput is obtained for delay-limited transmissions.
	\item The paper also provides an expression of overall energy efficiency for the considered system. Moreover, the critical value of spectrum sharing factor is deduced for which the OP of primary links with the proposed scheme exhibits the same OP as of direct communications (without spectrum sharing).
   \item  To evaluate the delay performance, the paper formulates an expression for the average end-to-end transmission time of the primary system.
	\item The paper reveals impacts of different system and channel parameters through extensive numerical and simulation results. The obtained results help to address some key physical-layer design aspects for practical deployments of such complex systems.
\end{itemize}

The rest of the paper is organized as follows. In Section \ref{sysmod}, the system model and proposed scheme are described. Specifically, Sections \ref{EH} and \ref{IP} present the considered EH model and IP signaling, respectively, and derive end-to-end instantaneous signal-to-noise ratios (SNRs). For evaluating the system performance, expressions of OP, system throughput, and energy efficiency are obtained in Section \ref{performance}. Numerical and simulation results are provided and discussed in Section \ref{numerical}. Finally, Section \ref{conclusion} concludes the paper.

\textit{Notations}: Throughout this paper, $f_{X}(\cdot)$ and $F_{X}(\cdot)$ represent the probability density function (PDF) and the cumulative distribution function (CDF) of a random variable $X$, respectively, and $\textmd{Pr}[\cdot]$ denotes probability. $\Gamma[\cdot,\cdot]$, $\Upsilon[\cdot,\cdot]$, and $\Gamma[\cdot]$ represent, respectively, the upper incomplete, the lower incomplete, and the complete gamma functions \cite[eq. (8.350)]{math}. $\mathds{E}[\cdot]$ and $\mathcal{K}_{v}(\cdot)$ denote expectation operation and $v$th order modified Bessel function of second kind  \cite[eq. (8.432.1)]{math}, respectively, whereas $\mathcal{W}_{u,v}(\cdot)$ represents Whittaker function  \cite[eq. (9.222)]{math}. Table \ref{table:notations} lists the
fundamental notations and parameters.

\section{System and Scheme Description}\label{sysmod}
\begin{figure}[t]
\centering
\includegraphics[width=3.65in]{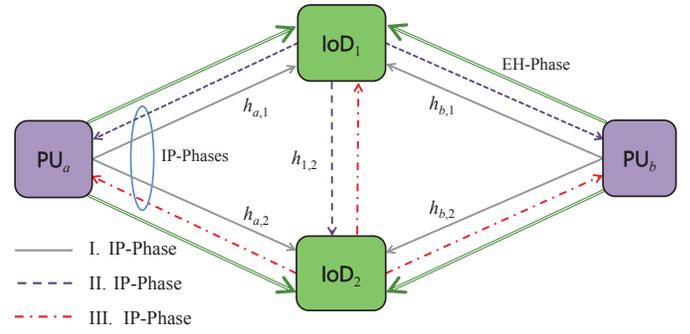}
\caption{System model for SWIPT-enabled bidirectional cognitive radio network.}
\label{fig1}
\end{figure}
Fig. \ref{fig1} depicts a SWIPT-enabled cognitive radio system considered in this paper. Two primary nodes ${\sf PU}_{a}$ and ${\sf PU}_{b}$ want to communicate to each other, but due to heavy shadowing or blockage, the direct link between them is not good enough to achieve specified target rates. A pair of proximate IoT devices\footnote{One can consider a generalized scenario of the considered system by assuming the presence of several potential pairs of IoT devices. Amongst them, the best pair can be selected through some appropriate selection process (see \cite{YPei2013} and references therein).}, referred to as ${\sf IoD}_{1}$ and ${\sf IoD}_{2}$, provides relay assistance to the primary transmissions and gets the opportunity to realize its bidirectional communications over the same licensed band. All the participating nodes (primary and IoT) operate in half-duplex mode and are equipped with single antenna devices.

The EH and IP processes can be done in separately dedicated time slots. To this end, one block duration is divided into two phases, i.e., EH phase and IP phase. During the EH phase, both IoDs harvest energy from the RF signals and store this energy to utilize it for providing relaying to the primary system and for their own transmissions. After EH phase, one round of end-to-end information exchanges between two PUs and two IoDs takes three IP phases. In the first IP phase or multiple access channel (MAC) phase, both ${\sf PU}_{a}$ and ${\sf PU}_{b}$ transmit their signals to both ${\sf IoD}_{1}$ and ${\sf IoD}_{2}$. After that, both ${\sf IoD}_{1}$ and ${\sf IoD}_{2}$ first decode the primary signals. On successful decoding, they apply bit-wise $\footnotesize{\textmd{XOR}}$ operation to generate a re-encoded primary signal for performing DF operation. Among ${\sf IoD}_{1}$ and ${\sf IoD}_{2}$, the first relaying IoD is the one who wants to communicate first to the other one. In the second IP phase or the first broadcast channel (BC) phase, the first relaying IoD broadcasts the encoded primary signal after adding its own signal intended for the other IoD. Likewise, in the third IP phase or the second BC phase, the other IoD applies the same procedure as done by the first one. If any IoD fails to decode both primary signals in the first IP phase, it transmits one-bit negative acknowledgment in the respective BC phase. At the receiving PUs, the SC scheme is employed to make use of two intended signal copies.

A block fading scenario is considered in this paper, where channel gains remain unchanged for one block duration. The channel gains of the links from ${\sf PU}_{ j }$ to ${\sf IoD}_{i}$ and from ${\sf IoD}_{i}$ to ${\sf PU}_{\hat{j}}$ are denoted as ${h}_{ j , i }$ and ${h}_{ i ,\hat{j}}$, respectively, for $ i \in\{1,2\}$ and $ j ,\hat{j}\in\{a,b\}$, where $ j \neq\hat{j}$. Likewise, the channel gain of the link from ${\sf IoD}_{i}$ to ${\sf IoD}_{\hat{i}}$ is denoted by ${h}_{ i ,\hat{i}}$. All the channel gains of individual hops are assumed to obey reciprocity, i.e., ${h}_{ i ,\hat{j}}={h}_{\hat{j}, i }$ and ${h}_{ i ,\hat{i}}={h}_{\hat{i}, i }$. Further, $h_{ j , i }$ for $ i \in\{1,2\}$ and $ j \in\{a,b\}$ follows Nakagami-$m$ distribution with fading severity $m_{ i  j }$ and average power $\Omega_{ i  j }$. The integer-valued fading parameters are adopted for modeling Nakagami-$m$ channels, through which a wide variety of wireless fading scenarios can be represented. It is also assumed that all the receiving terminals are affected by the additive white Gaussian noise (AWGN) with zero mean and variance $\sigma^{2}$.

\subsection{Energy Harvesting}\label{EH}
\begin{figure}[t]
\centering
\includegraphics[width=3.4in]{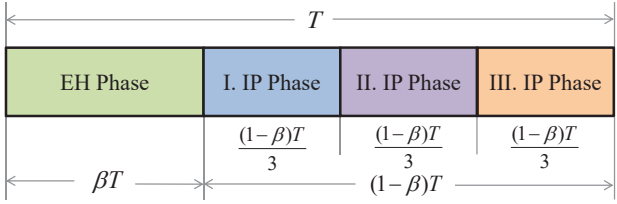}
\caption{Frame structure of TS-based SWIPT in the proposed cognitive radio network.}
\label{fig2}
\end{figure}

In TS-based EH approach as adopted in \cite{Nasir} and \cite{LiuIET}, one time slot is dedicated for harvesting energy from the RF signals and another slot for processing the information. In this paper, one transmission block duration $T$ is divided into two slots of durations $\beta T$ and $(1-\beta) T$ as shown in Fig. \ref{fig2}, where $0<\beta<1$. Herein, $\beta T$ is allocated for harvesting energy, whereas $(1-\beta) T$ is dedicated for information exchanges of primary and IoT systems. The value of $\beta$ that reflects the amount of harvested energy at the IoDs has a strong influence on the system performance in terms of achievable throughput and link reliability. The IP phase is further divided into three equal time slots, i.e., one MAC phase and two BC phases.

In the EH phase, the harvested energy at ${\sf IoD}_{ i }$ can be formulated as \cite{LiuIET}
\begin{align}\label{ehhh}
E_{ i }=\eta_{ i }\beta T(P_{a}|h_{a, i }|^{2}+P_{b}|h_{b, i }|^{2})
\end{align}
for $ i \in\{1,2\}$, where $0<\eta_{ i }<1$ represents the energy conversion efficiency of the EH circuit at ${\sf IoD}_{i}$, and $P_{a}\,\&\,P_{b}$ are transmit powers at ${\sf PU}_{a}\,\&\,{\sf PU}_{b}$, respectively. By using \eqref{ehhh}, the transmit power at ${\sf IoD}_{i}$ can be expressed as
\begin{align}\label{power}
P_{ i }=\frac{3\eta_{ i }\beta}{1-\beta}(P_{a}|h_{a, i }|^{2}+P_{b}|h_{b, i }|^{2}).
\end{align}
Without loss of generality, this paper assumes that all the harvested energy will be used for broadcasting the information signals at IoDs \cite{Nasir2}.
\subsection{Information Processing}\label{IP}
After the EH phase,  ${\sf PU}_{a}$ and ${\sf PU}_{b}$ transmit unit-energy symbols $x_{a}$ and $x_{b}$ in the first IP phase (MAC phase), respectively. Thereby, the received signals at ${\sf IoD}_{1}$ and ${\sf IoD}_{2}$ can be expressed, respectively, as
\begin{align}\label{sadjlk}
y_{1}=\sqrt{P_{a}}h_{a,1}x_{a}+\sqrt{P_{b}}h_{b,1}x_{b}+n_{1}
\end{align}
and
\begin{align}\label{sadj}
y_{2}=\sqrt{P_{a}}h_{a,2}x_{a}+\sqrt{P_{b}}h_{b,2}x_{b}+n_{2}
\end{align}
where $n_{1}\sim\mathcal{CN}(0,\sigma^{2})$ and $n_{2}\sim\mathcal{CN}(0,\sigma^{2})$ are AWGN components at ${\sf IoD}_{1}$ and ${\sf IoD}_{2}$, respectively.
After receiving the concurrent primary transmissions, both IoDs first decode $x_{a}$ and $x_{b}$ and then broadcast the combined primary signal using the DF operation.
\subsubsection{Decode-and-Forward Operation}\label{DF}
The IoDs can perform DF operation only when they successfully decode both the primary signals in the first IP phase. After decoding $x_{a}$ and $x_{b}$, the IoDs obtain a re-encoded symbol by performing bit-wise $\footnotesize{\textmd{XOR}}$ operation $(x_{a}\oplus x_{b})$ and utilize it for further transmissions. As in some practical applications, e.g., video streaming, gaming, and file transfer, the required data rates may be asymmetric for two opposite traffic flows. Therefore, the bit sequences corresponding to the primary signals may have different lengths. For ensuring the same bit sequence length, zero padding can be done on the shorter sequence \cite{RWang}.

Let ${\sf IoD}_{i}$ be the first relaying node and it broadcasts the symbol $x_{a}\oplus x_{b}$ by adding its own symbol $x_{i}$ intended for the other IoD. If $\mu_{i}$ represents the power splitting factor (resource allocation factor for primary transmissions) at ${\sf IoD}_{i}$, then the signal transmitted from ${\sf IoD}_{i}$ in the second IP phase (first BC phase) can be expressed as
\begin{align}\label{dff}
x^{\textmd{\tiny{BC}}}_{i}=\sqrt{\mu_{i} P_{i}}(x_{a}\oplus x_{b})+\sqrt{(1-\mu_{i}) P_{i}}x_{i}
\end{align}
where $P_{i}$ is the transmit power at ${\sf IoD}_{i}$. Further, the received signal at  ${\sf PU}_{j}$ in the first BC phase can be given as
\begin{align}\label{sdad}
y_{i,j}=\sqrt{\mu_{i} P_{i}}h_{i,j}(x_{a}\oplus x_{b})+\sqrt{(1-\mu_{i}) P_{i}}h_{i,j}x_{i}+n_{j}
\end{align}
where $n_{j}$ is AWGN variable at ${\sf PU}_{j}$. As both PUs know their respective transmitted signals, they can extract the desired information from the combined primary signal.
\subsubsection{End-to-End Instantaneous SNRs}\label{e2e}
Considering the IoT signal (interference to PUs) as additional noise and invoking the expression of $P_{i}$ from \eqref{power} into \eqref{sdad}, the end-to-end instantaneous SNR at ${\sf PU}_{j}$ can be expressed as
\begin{align}\label{lajad}
\gamma_{i,j}=\frac{\mu_{i}\zeta_{ij}|h_{i,j}|^{4}+\mu_{i}\zeta_{i\hat{j}}|h_{i,j}|^{2}|h_{i,\hat{j}}|^{2}}
{(1-\mu_{i})\zeta_{ij}|h_{i,j}|^{4}+(1-\mu_{i})\zeta_{i\hat{j}}|h_{i,j}|^{2}|h_{i,\hat{j}}|^{2}+1}
\end{align}
where $\zeta_{ij}=\frac{3\eta_{i}\rho_{j}\beta}{1-\beta}$ and $\zeta_{i\hat{j}}=\frac{3\eta_{i}\rho_{\hat{j}}\beta}{1-\beta}$ with $\rho_{j}=P_{j}/\sigma^{2}$ and $\rho_{\hat{j}}=P_{\hat{j}}/\sigma^{2}$, for $i\in\{1,2\}$, $j,\hat{j}\in\{a,b\}$, $j\neq\hat{j}$.
On the other hand, the received signal at ${\sf IoD}_{\hat{i}}$ in the second IP phase can be expressed as
\begin{align}\label{sdadad}
y_{i,\hat{i}}=\sqrt{\mu_{i} P_{i}}h_{i,\hat{i}}(x_{a}\oplus x_{b})+\sqrt{(1-\mu_{i}) P_{i}}h_{i,\hat{i}}x_{i}+n_{\hat{i}}
\end{align}
where $n_{\hat{i}}$ is AWGN variable at ${\sf IoD}_{\hat{i}}$. Since both IoDs can have the knowledge of the primary signals after decoding $x_{a}$ and $x_{b}$, they can remove the primary interference from the received signal. Thereby, the effective instantaneous SNR at ${\sf IoD}_{\hat{i}}$ in the first BC phase can be given as
\begin{align}\label{jjjg}
\gamma_{i,\hat{i}}=(1-\mu_{i})|h_{i,\hat{i}}|^{2}\left(\zeta_{ij}|h_{i,j}|^{2}+\zeta_{i\hat{j}}|h_{i,\hat{j}}|^{2}\right).
\end{align}

Similar to the second IP phase, ${\sf IoD}_{\hat{i}}$ also broadcasts the combined primary signal $(x_{a}\oplus x_{b})$ in the third IP phase (second BC phase) by adding its own signal $x_{\hat{i}}$ intended for ${\sf IoD}_{i}$. Likewise, the effective instantaneous SNR at ${\sf PU}_{j}$ in the third IP phase can be given as
\begin{align}\label{lgajad}
\gamma_{\hat{i},j}=\frac{\mu_{\hat{i}}\zeta_{\hat{i}j}|h_{\hat{i},j}|^{4}+\mu_{\hat{i}}\zeta_{\hat{i}\hat{j}}|h_{\hat{i},j}|^{2}|h_{\hat{i},\hat{j}}|^{2}}
{(1-\mu_{\hat{i}})\zeta_{\hat{i}j}|h_{\hat{i},j}|^{4}+(1-\mu_{\hat{i}})\zeta_{\hat{i}\hat{j}}|h_{\hat{i},j}|^{2}|h_{\hat{i},\hat{j}}|^{2}+1}
\end{align}
where $\zeta_{\hat{i}j}=\frac{3\eta_{\hat{i}}\rho_{j}\beta}{1-\beta}$ and $\zeta_{\hat{i}\hat{j}}=\frac{3\eta_{\hat{i}}\rho_{\hat{j}}\beta}{1-\beta}$ with $\rho_{j}=P_{j}/\sigma^{2}$ and $\rho_{\hat{j}}=P_{\hat{j}}/\sigma^{2}$, for $\hat{i}\in\{1,2\}$, $j,\hat{j}\in\{a,b\}$, $j\neq\hat{j}$.
Following similar steps as applied to obtain \eqref{jjjg}, the effective instantaneous SNR at ${\sf IoD}_{i}$ can be give as
\begin{align}\label{jjjsg}
\gamma_{\hat{i},i}=(1-\mu_{\hat{i}})|h_{\hat{i},i}|^{2}\left(\zeta_{\hat{i}j}|h_{\hat{i},j}|^{2}+\zeta_{\hat{i}\hat{j}}|h_{\hat{i},\hat{j}}|^{2}\right).
\end{align}
\begin{table}[]
	\centering
	\caption{List of parameters and their physical meaning/expression, where $i, \hat{i}\in\{1,2\}$ for $i\neq \hat{i}$, and $j, \hat{j}\in\{a,b\}$ for $j\neq \hat{j}$.}
	\label{table:notations}
	\begin{tabular}{c|c}
	  		\hline\hline
	  		\multicolumn{1}{c|}{\bfseries Parameter} & \multicolumn{1}{c}{\bfseries Meaning/Expression} \\ \hline
	  		$\eta_{i}$                   &  Energy conversion efficiency at ${\sf IoD}_{i}$\\ \hline
	  		$\beta$                   &  Time-switching factor\\ \hline
	  		$P_{j}$                   &  Transmit powers at ${\sf PU}_{j}$ \\ \hline
	  		$x_{j}$, $x_{i}$ & Transmit symbols at ${\sf PU}_{j}$ and ${\sf IoD}_{i}$  \\ \hline
	  		$h_{i,j}$, $h_{j,i}$                & Channel coefficients between ${\sf PU}_{j}$ and ${\sf IoD}_{i}$\\ \hline
	  		$h_{i,\hat{i}}$ & Channel coefficients between ${\sf IoD}_{i}$ and ${\sf IoD}_{\hat{i}}$\\ \hline
	  		$n_{i}$, $n_{j}$                     & AWGNs at ${\sf IoD}_{i}$ and ${\sf PU}_{j}$ with variance $\sigma^{2}$ \\ \hline
	  		$\mu_{i}$                           & Power splitting factor at ${\sf IoD}_{i}$                     \\ \hline
	  		$m_{ij}$                            & Fading severity parameter of $h_{i,j}$                \\ \hline
	  		$\Omega_{ij}$                       & Average power of $h_{i,j}$                            \\ \hline
	  		$r_{i}$, $r_{\hat{i}}$                    & Target rates at IoDs            \\ \hline
	  		$r_{j}$, $r_{\hat{j}}$                    & Target rates at PUs           \\ \hline
	  		$\mathcal{R}_{i,\hat{i}}$,  $\mathcal{R}_{i,j}$  & Instantaneous rates at IoDs and PUs \\ \hline
	  		${\overline\gamma_{\hat{i}}}$      & Target SNR at ${\sf IoD}_{\hat{i}}$             \\ \hline
	  		$\rho_{j}$                         &$P_{j}/\sigma^2$                        \\ \hline
	  		$\rho_{\hat{j}}$                         &$P_{\hat{j}}/\sigma^2$                      \\ \hline
	  		$\Theta_{j}$                        &$\rho_{j}/\rho_{\hat{j}}$               \\ \hline
	  		 $\varphi_{j}$                    &$2^{\frac{3 r_{j}}{1-\beta}}$                  \\ \hline
	  		 $\varphi_{\hat{j}}$             &$2^{\frac{3 r_{\hat{j}}}{1-\beta}}$             \\ \hline
	  		 $\mathcal{C}_{i}$      &$(\varphi_{j}\varphi_{\hat{j}}-\varphi_{\hat{j}})/\rho_{j}$ \\ \hline
	  		 $\mathcal{D}_{i}$ &    $(\varphi_{j}\varphi_{\hat{j}}-1)/ \rho_{\hat{j}}$ \\\hline
	  		 $\phi_{i}$ & $\mu_{i}-(1-\mu_{i})\gamma_{\textmd{th}}$ \\ \hline
	  		 $\gamma_{\textmd{th}}$ & $2^{\frac{3r_{\textmd{th}}}{1-\beta}}-1$ \\ \hline
	  		 $\widetilde{\gamma}_{j}$ & $2^{2r_{j}}-1$ \\ \hline
	  		 \hline
	  	\end{tabular}
\end{table}

\section{Performance Analysis}\label{performance}
This section first obtains closed-form expressions for user OP of primary and IoT systems under Nakagami-$m$ fading environment. Using these OP results, expressions of system throughput and energy efficiency are then provided for the considered system.

\subsection{Outage Probability of Primary System}
The OP is an important performance metric to quantify the link reliability of a wireless system over fading channels. With the proposed scheme, PUs can have two copies of their intended signals received from two IoDs in consecutive IP phases. Consequently, the outage event takes place at any PU if its instantaneous data rate achieved by exploiting both signal copies falls below a predefined target data rate. Mathematically, the user OP for the primary system can be computed as \cite{Pandhari}
\begin{align}\label{oopc}
\mathcal{P}_{\textmd{out},j}&=\textmd{Pr}[\mathcal{Q}_{i}]\,\textmd{Pr}[\mathcal{Q}_{\hat{i}}]\,\textmd{Pr}[\mathcal{R}_{{\rm sc},j}<r_{\textmd{th}}]
\nonumber\\
&+\textmd{Pr}[\mathcal{Q}_{i}]\,\left(1-\textmd{Pr}[\mathcal{Q}_{\hat{i}}]\right)\,\textmd{Pr}[\mathcal{R}_{i,j}<r_{\textmd{th}}]\nonumber\\
&+\left(1-\textmd{Pr}[\mathcal{Q}_{i}]\right)\,\textmd{Pr}[\mathcal{Q}_{\hat{i}}]\,\textmd{Pr}[\mathcal{R}_{\hat{i},j}<r_{\textmd{th}}]\nonumber\\
&+\left(1-\textmd{Pr}[\mathcal{Q}_{i}]\right)\,(1-\textmd{Pr}[\mathcal{Q}_{\hat{i}}])
\end{align}
for $j\in\{a,b\}$, $i,\hat{i}\in\{1,2\}$, $i\neq\hat{i}$. Here, $\textmd{Pr}[\mathcal{Q}_{i}]$ denotes the probability of successful decoding of $x_{a}$ and $x_{b}$ in the first IP phase at ${\sf IoD}_{i}$. In \eqref{oopc}, the first term accounts for the case when both IoDs successfully decode both signals ($x_{a}$ and $x_{b}$). On the other hand, the second and third terms represent the cases when only one IoD decodes the primary signals successfully. The forth term corresponds to case when both IoDs fail to decode the primary signals. Furthermore, $r_{\textmd{th}}=\max(r_{a}, r_{b})$, where $r_{a}$ and  $r_{b}$ denote the target data rates at ${\sf PU}_{a}$ and ${\sf PU}_{b}$, respectively. When the SC technique is employed at PUs to select the best signal copy (based on the maximum SNR), the instantaneous data rate can be given as
$\mathcal{R}_{{\rm sc},j}=\frac{1-\beta}{3}\log_{2}\left(1+\max\left(\gamma_{i,j},\gamma_{\hat{i},j}\right)\right)$.
Similarly, the instantaneous data rate related to individual signal copy at PUs can be expressed as
$\mathcal{R}_{i,j}=\frac{1-\beta}{3}\log_{2}\left(1+\gamma_{i,j}\right)$. In the first IP phase, a non-orthogonal multiple access scenario is considered, where both PUs transmit their signals $x_{a}$ and $x_{b}$ to IoDs over the same frequency band \cite{Popovski}. As such, the expression for correct decoding of both primary signals at ${\sf IoD}_{i}$ is provided in Lemma \ref{lem1} by following the procedure described in \cite{cover1} for decoding of simultaneously received signals.
\newtheorem{lemma}{Lemma}
\begin{lemma}\label{lem1}
The expression of $\textmd{Pr}[\mathcal{Q}_{i}]$, for $i\in\{1,2\}$, in (\ref{oopc}) can be given as
\begin{align}\label{saljd}
\textmd{Pr}[\mathcal{Q}_{i}] &=\left\{ \begin{array}{l}
\mathcal{P}_{Q_{i}},\quad\,\quad\,\textmd{for}\,\,\frac{m_{ij}}{\Omega_{ij} \rho_{j}}\neq\frac{m_{i\hat{j}}}{\Omega_{i\hat{j}} \rho_{\hat{j}}}\\
\widetilde{\mathcal{P}}_{Q_{i}},\quad\,\quad\,\textmd{for}\,\,\frac{m_{ij}}{\Omega_{ij} \rho_{j}}=\frac{m_{i\hat{j}}}{\Omega_{i\hat{j}} \rho_{\hat{j}}}
  \end{array}\right.
\end{align}
where $\mathcal{P}_{Q_{i}}$ and $\widetilde{\mathcal{P}}_{Q_{i}}$ are given by (\ref{sdlks}) and (\ref{ssdlks}) at the top of the next page, with $\mathcal{C}_{i}=(\varphi_{j}\varphi_{\hat{j}}-\varphi_{\hat{j}})/\rho_{j}$, $\mathcal{D}_{i}=(\varphi_{j}\varphi_{\hat{j}}-1)/ \rho_{\hat{j}}$,  $\varphi_{j}=2^{\frac{3 r_{j}}{1-\beta}}$ and $\varphi_{\hat{j}}=2^{\frac{3 r_{\hat{j}}}{1-\beta}}$.
\end{lemma}

\begin{figure*}[!t]
\begin{align}\label{sdlks}
\mathcal{P}_{Q_{i}}&=1-\frac{\Upsilon\left[m_{i\hat{j}},\frac{m_{i\hat{j}}}{\Omega_{i\hat{j}}}\left(\frac{\varphi_{j}-1}{ \rho_{\hat{j}}}\right)\right]}
{\Gamma[m_{i\hat{j}}]}
\left(1-\frac{\Upsilon\left[m_{ij},\frac{m_{ij}}{\Omega_{ij}}\mathcal{C}_{i}\right]}{\Gamma[m_{ij}]}\right)-
\frac{\Upsilon\left[m_{ij},\frac{m_{ij}}{\Omega_{ij}}\mathcal{C}_{i}\right]}{\Gamma[m_{ij}]}
+\frac{\left(\frac{m_{ij}}{\Omega_{ij}}\right)^{m_{ij}}}{\Gamma[m_{ij}]}{\mathrm{e}}^{-\frac{m_{i\hat{j}}}{\Omega_{i\hat{j}}}\mathcal{D}_{i}}
\sum^{m_{i\hat{j}}-1}_{k=0}\frac{\left(\frac{m_{i\hat{j}}}{\Omega_{i\hat{j}}}\right)^{k}}{k!}\nonumber\\
&\times\sum^{k}_{q=0}\binom{k}{q}\mathcal{D}_{i}^{q}(-\Theta_{j})^{k-q}\sum_{p=0}^{m_{ij}+k-q-1}\frac{\binom{m_{ij}+k-q-1}{p}(-1)^{p}p!}
{\left(\frac{\rho_{j}m_{i\hat{j}}}{\rho_{\hat{j}}\Omega_{i\hat{j}}}-\frac{m_{ij}}{\Omega_{ij}}\right)^{p+1}}
\bigg({\mathrm{e}}^{\left(\frac{\rho_{j}m_{i\hat{j}}}{\rho_{\hat{j}}\Omega_{i\hat{j}}}-\frac{m_{ij}}{\Omega_{ij}}\right)\mathcal{C}_{i}}
\mathcal{C}_{i}^{m_{ij}+k-q-1-p}
-{\mathrm{e}}^{\left(\frac{\rho_{j}m_{i\hat{j}}}{\rho_{\hat{j}}\Omega_{i\hat{j}}}-\frac{m_{ij}}{\Omega_{ij}}\right)\left(\frac{\varphi_{\hat{j}}-1}{ \rho_{j}}\right)}\nonumber\\
&\times\left(\frac{\varphi_{\hat{j}}-1}{ \rho_{j}}\right)^{m_{ij}+k-q-1-p}\bigg).
\end{align}
\hrulefill
\end{figure*}

\begin{figure*}[!t]
\begin{align}\label{ssdlks}
\widetilde{\mathcal{P}}_{Q_{i}}&=1-\frac{\Upsilon\left[m_{i\hat{j}},\frac{m_{i\hat{j}}}{\Omega_{i\hat{j}}}\left(\frac{\varphi_{j}-1}{ \rho_{\hat{j}}}\right)\right]}
{\Gamma[m_{i\hat{j}}]}
\left(1-\frac{\Upsilon\left[m_{ij},\frac{m_{ij}}{\Omega_{ij}}\mathcal{C}_{i}\right]}{\Gamma[m_{ij}]}\right)-
\frac{\Upsilon\left[m_{ij},\frac{m_{ij}}{\Omega_{ij}}\mathcal{C}_{i}\right]}{\Gamma[m_{ij}]}
+\frac{\left(\frac{m_{ij}}{\Omega_{ij}}\right)^{m_{ij}}}{\Gamma[m_{ij}]}{\mathrm{e}}^{-\frac{m_{i\hat{j}}}{\Omega_{i\hat{j}}}\mathcal{D}_{i}}\nonumber\\
&\times\sum^{m_{i\hat{j}}-1}_{k=0}\frac{\left(\frac{m_{i\hat{j}}}{\Omega_{i\hat{j}}}\right)^{k}}{k!}
\sum^{k}_{q=0}\binom{k}{q}\frac{\mathcal{D}_{i}^{q}(-\Theta_{j})^{k-q}}{m_{ij}+k-q}\left(\mathcal{C}_{i}^{m_{ij}+k-q}-\left(\frac{\varphi_{\hat{j}}-1}{ \rho_{j}}\right)^{m_{ij}+k-q}\right).
\end{align}
\hrulefill
\end{figure*}
\begin{IEEEproof}
Consider $Y_{i}\triangleq|h_{j,i}|^{2}$ and $Z_{i}\triangleq|h_{\hat{j},i}|^{2}$ for $i\in\{1,2\}$, $j,\hat{j}\in\{b,m\}$ with $j\neq\hat{j}$, which are Gamma-distributed random variables with PDFs $f_{Y_{i}}(y_{i})=\left(\frac{m_{ij}}{\Omega_{ij}}\right)^{m_{ij}} \frac{y^{m_{ij}-1}_{i}}{\Gamma[m_{ij}]}{\mathrm e}^{-\frac{m_{ij}y_{i}}{\Omega_{ij}}},\,y_{i}\geq0$, and $f_{Z_{i}}(z_{i})=\left(\frac{m_{i\hat{j}}}{\Omega_{i\hat{j}}}\right)^{m_{i\hat{j}}} \frac{z^{m_{i\hat{j}}-1}_{i}}{\Gamma[m_{i\hat{j}}]}{\mathrm e}^{-\frac{m_{i\hat{j}}z_{i}}{\Omega_{i\hat{j}}}},\,z_{i}\geq0$. In the first IP phase, for decoding the primary signals at ${\sf IoD}_{i}$, the following three conditions should be satisfied \cite{RWang}, \cite{Zhang2015}
\begin{align}\label{sasljd}
\left\{ \begin{array}{l}
\frac{1-\beta}{3}\log_{2}(1+ \rho_{j}Y_{i})\geq r_{\hat{j}}\\
\frac{1-\beta}{3}\log_{2}(1+ \rho_{\hat{j}}Z_{i})\geq r_{j}\\
\frac{1-\beta}{3}\log_{2}(1+ \rho_{j}Y_{i}+ \rho_{\hat{j}}Z_{i})\geq r_{\hat{j}}+r_{j}
\end{array}\right.
\end{align}
where $\rho_{j}=P_{j}/\sigma^{2}$. Based on (\ref{sasljd}), one can formulate the expression of  $\textmd{Pr}[\mathcal{Q}_{i}]$  as
\begin{align}\label{hfhtt}
\textmd{Pr}[\mathcal{Q}_{i}]&=\int^{\infty}_{\frac{\varphi_{\hat{j}}-1}{ \rho_{j}}}  f_{Y_{i}}(y_{i})\int^{\infty}_{\frac{\varphi_{j}-1}{ \rho_{\hat{j}}}}
f_{Z_{i}}(z_{i}) dz_{i} dy_{i}\nonumber\\
&-\int^{\mathcal{C}_{i}}_{\frac{\varphi_{\hat{j}}-1}{ \rho_{j}}}     f_{Y_{i}}(y_{i})
\int^{\mathcal{D}_{i}-\Theta_{j} y_{i}}_{\frac{\varphi_{j}-1}{ \rho_{\hat{j}}}}
f_{Z_{i}}(z_{i}) dz_{i} dy_{i}
\end{align}
where $\mathcal{C}_{i}$ and $\mathcal{D}_{i}$ are defined after \eqref{saljd} and $\Theta_{j}=\rho_{j}/\rho_{\hat{j}}$. On rearranging the limits of \eqref{hfhtt}, $\textmd{Pr}[\mathcal{S}_{i}]$ can be represented as
\begin{align}\label{hfhtttz}
\textmd{Pr}[\mathcal{Q}_{i}]&=\int^{\infty}_{\frac{\varphi_{\hat{j}}-1}{ \rho_{j}}}  f_{Y_{i}}(y_{i})\int^{\infty}_{\frac{\varphi_{j}-1}{ \rho_{\hat{j}}}} f_{Z_{i}}(z_{i}) dz_{i} dy_{i}\nonumber\\
&-\int^{\mathcal{C}_{i}}_{\frac{\varphi_{\hat{j}}-1}{ \rho_{j}}}f_{Y_{i}}(y_{i})
\int^{\mathcal{D}_{i}-\Theta_{j} y_{i}}_{0} f_{Z_{i}}(z_{i}) dz_{i} dy_{i}\nonumber\\
&+\int^{\mathcal{C}_{i}}_{\frac{\varphi_{\hat{j}}-1}{ \rho_{j}}} f_{Y_{i}}(y_{i})
\int^{\frac{\varphi_{j}-1}{ \rho_{\hat{j}}}}_{0} f_{Z_{i}}(z_{i}) dz_{i} dy_{i}.
\end{align}
By inserting the respective PDFs in \eqref{hfhtttz} and utilizing \cite[eq. 3.381]{math}, one can express it as
\begin{align}\label{yjyyyi}
\mathcal{P}_{Q_{i}}&=1-\frac{\Upsilon\left[m_{i\hat{j}},\frac{m_{i\hat{j}}}{\Omega_{i\hat{j}}}\left(\frac{\varphi_{j}-1}{ \rho_{\hat{j}}}\right)\right]}
{\Gamma[m_{i\hat{j}}]}
\left(1-\frac{\Upsilon\left[m_{ij},\frac{m_{ij}}{\Omega_{ij}}\mathcal{C}_{i}\right]}{\Gamma[m_{ij}]}\right)\nonumber\\
&-\frac{\Upsilon\left[m_{ij},\frac{m_{ij}}{\Omega_{ij}}\mathcal{C}_{i}\right]}{\Gamma[m_{ij}]}
+\frac{\left(\frac{m_{ij}}{\Omega_{ij}}\right)^{m_{ij}}}{\Gamma[m_{ij}]}
{\mathrm e}^{-\frac{m_{i\hat{j}}}{\Omega_{i\hat{j}}}\mathcal{D}_{i}}\nonumber\\
&\times\sum^{m_{i\hat{j}}-1}_{k=0}\frac{\left(\frac{m_{i\hat{j}}}{\Omega_{i\hat{j}}}\right)^{k}}{k!}
\sum^{k}_{q=0}\binom{k}{q}\mathcal{D}_{i}^{q}(-\Theta_{j})^{k-q}\nonumber\\
&\times\int^{\mathcal{C}_{i}}_{\frac{\varphi_{\hat{j}}-1}{ \rho_{j}}}y^{m_{ij}-1+k-q}_{i}{\mathrm e}^{-\left(\frac{m_{ij}}{\Omega_{ij}}-\frac{\Theta_{j}m_{i\hat{j}}}{\Omega_{i\hat{j}}}\right)y_{i}}dy_{i}.
\end{align}
On solving the last integration of \eqref{yjyyyi} with the help of \cite[eq. 2.321]{math}, the expression of $\textmd{Pr}[\mathcal{Q}_{i}]$ can be given as in (\ref{saljd}).
\end{IEEEproof}
Next, the probability term $\mathcal{P}_{{\rm sc},j}\triangleq \textmd{Pr}[\mathcal{R}_{{\rm sc},j}<r_{\textmd{th}}]$ in \eqref{oopc} can be formulated as
\begin{align}\label{ssdawe}
\mathcal{P}_{{\rm sc},j}&=\textmd{Pr}[\max(\gamma_{i,j},\gamma_{\hat{i},j})<\gamma_{\textmd{th}}]
=\prod^{2}_{i=1}F_{\gamma_{i,j}}(\gamma_{\textmd{th}})
\end{align}
where $\gamma_{\textmd{th}}=2^{\frac{3r_{\textmd{th}}}{1-\beta}}-1$. Further, the expression of $F_{\gamma_{i,j}}(\gamma_{\textmd{th}})$ is given in the following Lemma.

\newtheorem{lemmaa}{Lemma}
\begin{lemma}\label{lem2}
The CDF $F_{\gamma_{i,j}}(\gamma_{\textmd{th}})$ can be expressed as
\begin{align}\label{slkkdl}
F_{\gamma_{i,j}}(\gamma_{\textmd{th}})&= \left\{   \begin{array}{l}
\widetilde{F}_{\gamma_{i,j}}\quad\textmd{for}\,\, \frac{\gamma_{\textmd{th}}}{1+\gamma_{\textmd{th}}} < \mu_{i} < 1\\
1\quad\,\,\,\quad\textmd{for}\,\,\,0 < \mu_{i} < \frac{\gamma_{\textmd{th}}}{1+\gamma_{\textmd{th}}}
  \end{array}\right.
\end{align}
where
\begin{align}\label{fffwef}
\widetilde{F}_{\gamma_{i,j}}&=\frac{\Upsilon\left[m_{ij},\frac{m_{ij}}{\Omega_{ij}}\sqrt{\frac{\gamma_{\textmd{th}}}{\zeta_{ij\phi_{i}}}}\right]}{\Gamma[m_{ij}]}
-\sum^{m_{i\hat{j}}-1}_{k=0}\frac{\left(\frac{m_{i\hat{j}}}{\Omega_{i\hat{j}}\phi_{i}\zeta_{i\hat{j}}}\right)^{k}}{k!}\nonumber\\
&\times\frac{\left(\frac{m_{ij}}{\Omega_{ij}}\right)^{m_{ij}}}{\Gamma[m_{ij}]}
\sum^{k}_{n=0}\binom{k}{n}\gamma_{\textmd{th}}^{n}(-\zeta_{ij}\phi_{i})^{k-n}I_{m},
\end{align}
and $\phi_{i}=\mu_{i}-(1-\mu_{i})\gamma_{\textmd{th}}$. Further, the expression of $I_{m}$ is given for two cases, as
\begin{align}\label{slkksdl}
I_{m}&=\left\{\!\! \begin{array}{l}
I_{1},\quad\textmd{for} \quad\frac{m_{i\hat{j}}\zeta_{ij}}{\Omega_{i\hat{j}}\zeta_{i\hat{j}}}\neq\frac{m_{ij}}{\Omega_{ij}}\\
I_{2},\quad\textmd{for}\quad \frac{m_{i\hat{j}}\zeta_{ij}}{\Omega_{i\hat{j}}\zeta_{i\hat{j}}}=\frac{m_{ij}}{\Omega_{ij}}
  \end{array}\right.
\end{align}
where $I_{1}$ and $I_{2}$ are given in \eqref{lslje} and \eqref{lsslje}  on the next page.
\begin{figure*}[t]
\begin{align}\label{lslje}
I_{1}&=\sum^{\infty}_{p=0}\frac{\left(\frac{m_{i\hat{j}}\zeta_{ij}}{\Omega_{i\hat{j}}\zeta_{i\hat{j}}}-\frac{m_{ij}}{\Omega_{ij}}\right)^{p}}{p!}
\left(\frac{m_{i\hat{j}}\gamma_{\textmd{th}}}{\Omega_{i\hat{j}}\phi_{i}\zeta_{i\hat{j}}}\right)^{p+k-2n+m_{ij}}
\left(\frac{m_{i\hat{j}}\gamma_{\textmd{th}}}{\Omega_{i\hat{j}}\phi_{i}\zeta_{i\hat{j}}}\sqrt{\frac{\zeta_{ij}\phi_{i}}{\gamma_{\textmd{th}}}}\right)^{
-\frac{(p+k-2n+m_{ij}+1)}{2}}{\mathrm{e}}^{-\frac{m_{i\hat{j}}\gamma_{\textmd{th}}}{2\Omega_{i\hat{j}}\phi_{i}\zeta_{i\hat{j}}}\sqrt{\frac{\zeta_{ij}\phi_{i}}{\gamma_{\textmd{th}}}}}\nonumber\\
&\times \mathcal{W}_{-\frac{(p+k-2n+m_{ij}+1)}{2},\frac{1-(p+k-2n+m_{ij}+1)}{2}}\left(\frac{m_{i\hat{j}}\gamma_{\textmd{th}}}{\Omega_{i\hat{j}}\phi_{i}\zeta_{i\hat{j}}}
\sqrt{\frac{\zeta_{ij}\phi_{i}}{\gamma_{\textmd{th}}}}\right).
\end{align}
\hrulefill
\end{figure*}
\begin{figure*}[t]
\begin{align}\label{lsslje}
I_{2}&=\left(\frac{m_{i\hat{j}}\gamma_{\textmd{th}}}{\Omega_{i\hat{j}}\phi_{i}\zeta_{i\hat{j}}}\right)^{k-2n+m_{ij}}
\left(\frac{m_{i\hat{j}}\gamma_{\textmd{th}}}{\Omega_{i\hat{j}}\phi_{i}\zeta_{i\hat{j}}}\sqrt{\frac{\zeta_{ij}\phi_{i}}{\gamma_{\textmd{th}}}}\right)^{
-\frac{(k-2n+m_{ij}+1)}{2}}{\mathrm{e}}^{-\frac{m_{i\hat{j}}\gamma_{\textmd{th}}}{2\Omega_{i\hat{j}}\phi_{i}\zeta_{i\hat{j}}}\sqrt{\frac{\zeta_{ij}\phi_{i}}{\gamma_{\textmd{th}}}}}\nonumber\\
&\times \mathcal{W}_{-\frac{(k-2n+m_{ij}+1)}{2},\frac{1-(k-2n+m_{ij}+1)}{2}}\left(\frac{m_{i\hat{j}}\gamma_{\textmd{th}}}{\Omega_{i\hat{j}}\phi_{i}\zeta_{i\hat{j}}}
\sqrt{\frac{\zeta_{ij}\phi_{i}}{\gamma_{\textmd{th}}}}\right).
\end{align}
\hrulefill
\end{figure*}
\begin{IEEEproof}
The CDF $F_{\gamma_{i,j}}(\gamma_{\textmd{th}})$ can be formulated using (\ref{lajad}) as
\begin{align}\label{luuhh}
F_{\gamma_{i,j}}(\gamma_{\textmd{th}})&=\textmd{Pr}\left[\frac{\mu_{i}\zeta_{ij}Y^{2}_{i}+\mu_{i}\zeta_{i\hat{j}}Y_{i}Z_{i}}
{(1-\mu_{i})(\zeta_{ij}Y^{2}_{i}+\zeta_{i\hat{j}}Y_{i}Z_{i})+1}<\gamma_{\textmd{th}}\right]\nonumber\\
&=\textmd{Pr}\left[Z_{i}<\frac{\left((1-\mu_{i})\gamma_{\textmd{th}}-\mu_{i}\right)\zeta_{ij}Y^{2}_{i}+\gamma_{\textmd{th}}}{
	\left(\mu_{i}-(1-\mu_{i})\gamma_{\textmd{th}}\right)\zeta_{i\hat{j}}Y_{i}}\right].
\end{align}
When the term $\left(\mu_{i}-(1-\mu_{i})\gamma_{\textmd{th}}\right)\leq0$, the CDF $F_{\gamma_{i,j}}(\gamma_{\textmd{th}})=1$. On the other hand, when $\left(\mu_{i}-(1-\mu_{i})\gamma_{\textmd{th}}\right)>0$, the expression of $F_{\gamma_{i,j}}(\gamma_{\textmd{th}})$ can be formulated as
\begin{align}\label{ljljio}
F_{\gamma_{i,j}}(\gamma_{\textmd{th}})&=\int^{\sqrt{\frac{\gamma_{\textmd{th}}}{\zeta_{ij}\phi_{i}}}}_{y_{i}=0}f_{Y_{i}}(y_{i})
\int^{\frac{\gamma_{\textmd{th}}-\zeta_{ij}\phi_{i}y^{2}_{i}}{\phi_{i}\zeta_{i\hat{j}}y_{i}}}_{z_{i}=0}f_{Z_{i}}(z_{i})dz_{i}\,dy_{i}
\end{align}
where $\phi_{i}$ is defined after \eqref{fffwef}. After inserting the respective PDFs in \eqref{ljljio} and applying \cite[eqs. 3.381.1, 8.352.1]{math}, one obtains
\begin{align}\label{fsjkl}
&F_{\gamma_{i,j}}(\gamma_{\textmd{th}})=\nonumber\\
&\frac{\Upsilon\left[m_{ij},\frac{m_{ij}}{\Omega_{ij}}\sqrt{\frac{\gamma_{\textmd{th}}}{\zeta_{ij}\phi_{i}}}\right]}{\Gamma[m_{ij}]}
-\sum^{m_{i\hat{j}}-1}_{k=0}\frac{\left(\frac{m_{i\hat{j}}}{\Omega_{i\hat{j}}\phi_{i}\zeta_{i\hat{j}}}\right)^{k}}{k!}\nonumber\\
&\times\frac{\left(\frac{m_{ij}}{\Omega_{ij}}\right)^{m_{ij}}}{\Gamma[m_{ij}]}
\sum^{k}_{n=0}\binom{k}{n}\gamma_{\textmd{th}}^{n}(-\zeta_{ij}\phi_{i})^{k-n}\nonumber\\
&\times\int^{\sqrt{\frac{\gamma_{\textmd{th}}}{\zeta_{ij}\phi_{i}}}}_{y_{i}=0}y^{m_{ij}+k-2n-1}_{i}{\mathrm e}^{-\frac{m_{i\hat{j}}\gamma_{\textmd{th}}}
	{\Omega_{i\hat{j}}\phi_{i}\zeta_{i\hat{j}}y_{i}}}{\mathrm e}^{\left(\frac{m_{i\hat{j}}\zeta_{ij}}{\Omega_{i\hat{j}}\zeta_{i\hat{j}}}
	-\frac{m_{ij}}{\Omega_{ij}}\right)y_{i}}dy_{i}.
\end{align}
As the solution of \eqref{fsjkl} for the general case is mathematically intractable, this paper provides the solutions for two cases.\\
\textit{Case-1:} For $\frac{m_{i\hat{j}}\zeta_{ij}}{\Omega_{i\hat{j}}\zeta_{i\hat{j}}}
\neq\frac{m_{ij}}{\Omega_{ij}}$, after applying Maclaurin series expansion for the last exponential term of \eqref{fsjkl}, one has
\begin{align}\label{fjkslsj}
\!\!\!\!F_{\gamma_{i,j}}(\gamma_{\textmd{th}})\!&=\!\frac{\Upsilon\left[m_{ij},\frac{m_{ij}}{\Omega_{ij}}\sqrt{\frac{\gamma_{\textmd{th}}}{\zeta_{ij}\phi_{i}}}\right]}{\Gamma[m_{ij}]}
\!-\!\!\sum^{m_{i\hat{j}}-1}_{k=0}\!\!\frac{\left(\frac{m_{i\hat{j}}}{\Omega_{i\hat{j}}\phi_{i}\zeta_{i\hat{j}}}\right)^{k}}{k!}
\frac{\left(\frac{m_{ij}}{\Omega_{ij}}\right)^{m_{ij}}}{\Gamma[m_{ij}]}\nonumber\\
&\times\sum^{k}_{n=0}\binom{k}{n}\gamma_{\textmd{th}}^{n}(-\zeta_{ij}\phi_{i})^{k-n}\sum^{\infty}_{p=0}
\frac{\left(\frac{m_{i\hat{j}}\zeta_{ij}}{\Omega_{i\hat{j}}\zeta_{i\hat{j}}}
	-\frac{m_{ij}}{\Omega_{ij}}\right)^{p}}{p!}\nonumber\\
&\times\int^{\sqrt{\frac{\gamma_{\textmd{th}}}{\zeta_{ij}\phi_{i}}}}_{y_{i}=0}y^{p+m_{ij}+k-2n-1}_{i}{\mathrm e}^{-\frac{m_{i\hat{j}}\gamma_{\textmd{th}}}
	{\Omega_{i\hat{j}}\phi_{i}\zeta_{i\hat{j}}y_{i}}}dy_{i}.
\end{align}
By applying change of variables with $t=\frac{m_{i\hat{j}}\gamma_{\textmd{th}}}
{\Omega_{i\hat{j}}\phi_{i}\zeta_{i\hat{j}}y_{i}}$, and then utilizing \cite[eq. 3.381.6]{math}, the solution is obtained as given in \eqref{slkkdl} and \eqref{lslje}.\\
\textit{Case-2:} For $\frac{m_{i\hat{j}}\zeta_{ij}}{\Omega_{i\hat{j}}\zeta_{i\hat{j}}}
=\frac{m_{ij}}{\Omega_{ij}}$, making change of variables  $t=\frac{m_{i\hat{j}}\gamma_{\textmd{th}}}
{\Omega_{i\hat{j}}\phi_{i}\zeta_{i\hat{j}}y_{i}}$ in \eqref{fsjkl}, and then using \cite[eq. 3.381.6]{math}, one obtains the solution as given in \eqref{slkkdl} and \eqref{lsslje}.
\end{IEEEproof}
\end{lemma}

The same derivation steps can be followed to obtain expressions of other probabilities of \eqref{oopc} by replacing $i$ with $\hat{i}$ and vice-versa in Lemma \ref{lem2}. On inserting \eqref{saljd}-\eqref{lsslje} into \eqref{oopc}, one can get the desired OP expression for the primary system.

\subsection{Outage Probability of IoT System}
An outage event occurs at the IoD if any IoDs fail to decode the primary signals or the instantaneous rate achieved at that node falls below a predefined target rate. Following this, the user OP of the IoT system can be formulated as
\begin{align}\label{oopd}
\mathcal{P}_{\textmd{out},\hat{i}}=1-\textmd{Pr}[\mathcal{Q}_{i}]\,\textmd{Pr}[\mathcal{Q}_{\hat{i}}]\textmd{Pr}[\mathcal{R}_{i,\hat{i}}>r_{\hat{i}}]
\end{align}
for $i,\hat{i}\in\{1,2\}$ and $i\neq\hat{i}$, where $r_{\hat{i}}$ is the target rate at ${\sf IoD}_{\hat{i}}$. Moreover, $\mathcal{R}_{i,\hat{i}}=((1-\beta)/3)\log_{2}(1+\gamma_{i,\hat{i}})$ is the instantaneous rate at ${\sf IoD}_{\hat{i}}$. The decoding probabilities are already derived in Lemma \ref{lem1} and the remaining term can be calculated as $\textmd{Pr}[\mathcal{R}_{i,\hat{i}}>r_{\hat{i}}]=1-\textmd{Pr}[\mathcal{R}_{i,\hat{i}}<r_{\hat{i}}]\triangleq 1-F_{\gamma_{i,\hat{i}}}(\overline{\gamma}_{\hat{i}})$. Further, the expression of $F_{\gamma_{i,\hat{i}}}(\overline{\gamma}_{\hat{i}})$ is given in the following Lemma.

\newtheorem{lemmaaa}{Lemma}
\begin{lemma}\label{lem3}
	The CDF $F_{\gamma_{i,\hat{i}}}({\overline{\gamma}_{\hat{i}}})$ can be expressed as
	\begin{align}\label{sioio}
	F_{\gamma_{i,\hat{i}}}(\overline{\gamma}_{\hat{i}})&= 1-\sum^{m_{ij}-1}_{q=0}\frac{\left(\frac{m_{ij}\overline{\gamma}_{\hat{i}}}{\Omega_{ij}(1-\mu_{i})\zeta_{ij}}\right)^{q}}{\Gamma[m_{i\hat{i}}]q!}
	\left(\frac{m_{i\hat{i}}}{\Omega_{i\hat{i}}}\right)^{m_{i\hat{i}}}\nonumber\\
	&\times 2\left(\frac{m_{ij}\overline{\gamma}_{\hat{i}}\Omega_{i\hat{i}}}{\Omega_{ij}m_{i\hat{i}}(1-\mu_{i})\zeta_{ij}}\right)^{\frac{m_{i\hat{i}}-q}{2}}\nonumber\\
	 &\times\mathcal{K}_{m_{i\hat{i}}-q}\left(2\sqrt{\frac{m_{ij}\overline{\gamma}_{\hat{i}}m_{i\hat{i}}}{\Omega_{ij}(1-\mu_{i})\zeta_{ij}\Omega_{i\hat{i}}}}\right)
	-I_{3}
	\end{align}
	where $\overline{\gamma}_{\hat{i}}=2^{\frac{3r_{\hat{i}}}{1-\beta}}-1$ and the expression of $I_{3}$ is given on top of the next page.
	\begin{figure*}[t]
	\begin{align}\label{lsae}
				 I_{3}&=\frac{\left(\frac{m_{i\hat{i}}}{\Omega_{i\hat{i}}}\right)^{m_{i\hat{i}}}}{\Gamma[m_{i\hat{i}}]}
				 \sum^{m_{i\hat{j}}-1}_{k=0}\frac{\left(\frac{m_{i\hat{j}}}{\Omega_{i\hat{j}}(1-\mu_{i})\zeta_{i\hat{j}}}\right)^{k}}{k!}
				 \sum^{k}_{s=0}\binom{k}{s}\overline{\gamma}_{\hat{i}}^{s}(-1)^{k-s}((1-\mu_{i})\zeta_{ij})^{k-s}\sum^{\infty}_{p=0}\frac{\left(\frac{m_{i\hat{j}}\zeta_{ij}}
					 {\Omega_{i\hat{j}}\zeta_{i\hat{j}}}\right)^{p}}{p!}\left(\frac{m_{ij}}{\Omega_{ij}}\right)^{-p-k+s} \nonumber\\
				 &\times2\Gamma[p+m_{ij}+k-s]\Bigg(\left(\frac{m_{i\hat{j}}\overline{\gamma}_{\hat{i}}\Omega_{i\hat{i}}}{\Omega_{i\hat{j}}(1-\mu_{i})
					\zeta_{i\hat{j}}m_{i\hat{i}}}\right)^{\frac{m_{i\hat{i}}-s}{2}}\mathcal{K}_{m_{i\hat{i}}-s}
				 \left(2\sqrt{\frac{m_{i\hat{j}}\overline{\gamma}_{\hat{i}}m_{i\hat{i}}}{\Omega_{i\hat{j}}(1-\mu_{i})\zeta_{i\hat{j}}\Omega_{i\hat{i}}}}\right)
				 -\sum^{p+m_{ij}+k-s-1}_{l=0}\frac{\left(\frac{m_{ij}\overline{\gamma}_{\hat{i}}}{\Omega_{ij}(1-\mu_{i})\zeta_{ij}}\right)^{l}}{l!}\nonumber\\
				 &\times\left(\left(\frac{m_{i\hat{j}}\overline{\gamma}_{\hat{i}}}{\Omega_{i\hat{j}}(1-\mu_{i})\zeta_{i\hat{j}}}+
				 \frac{m_{ij}\overline{\gamma}_{\hat{i}}}{\Omega_{ij}(1-\mu_{i})\zeta_{ij}}\right)\frac{\Omega_{i\hat{i}}}{m_{i\hat{i}}}\right)^{\frac{m_{i\hat{i}}-s-l}{2}}
				 \mathcal{K}_{m_{i\hat{i}}-s-l}\left(2\sqrt{\left(\frac{m_{i\hat{j}}\overline{\gamma}_{\hat{i}}}{\Omega_{i\hat{j}}(1-\mu_{i})\zeta_{i\hat{j}}}+
					 \frac{m_{ij}\overline{\gamma}_{\hat{i}}}{\Omega_{ij}(1-\mu_{i})\zeta_{ij}}\right)\frac{m_{i\hat{i}}}{\Omega_{i\hat{i}}}}\right)\Bigg).
				\end{align}
		\hrulefill
	\end{figure*}
	\begin{IEEEproof}
	$X\triangleq|h_{i,\hat{i}}|^{2}$ is a Gamma distributed random variable with PDF as $f_{X}(x)=\left(\frac{m_{i\hat{i}}}{\Omega_{i\hat{i}}}\right)^{m_{i\hat{i}}} \frac{x^{m_{i\hat{i}}-1}}{\Gamma[m_{i\hat{i}}]}{\mathrm e}^{-\frac{{m_{i\hat{i}}}x}{\Omega_{i\hat{i}}}},\,x\geq0$. On utilizing \eqref{jjjg}, the CDF $F_{\gamma_{i,\hat{i}}}(\overline{\gamma}_{\hat{i}})$ can be formulated as
\begin{align}\label{jjjssg}
F_{\gamma_{i,\hat{i}}}(\overline{\gamma}_{\hat{i}})&=\textmd{Pr}\left[(1-\mu_{i})\zeta_{ij}
Y_{i}X+(1-\mu_{i})\zeta_{i\hat{j}}Z_{i}X<\overline{\gamma}_{\hat{i}}\right]\nonumber\\
&=\textmd{Pr}\left[Z_{i}<\frac{\overline{\gamma}_{\hat{i}}-(1-\mu_{i})\zeta_{ij}Y_{i}X}{(1-\mu_{i})\zeta_{i\hat{j}}X}\right].
\end{align}
Based on \eqref{jjjssg}, $F_{\gamma_{i,\hat{i}}}(\overline{\gamma}_{\hat{i}})$ can be formulated in integration form as
\begin{align}\label{kmklw}
F_{\gamma_{i,\hat{i}}}(\overline{\gamma}_{\hat{i}})&=\int^{\infty}_{x=0}f_{X}(x)\int^{\frac{\overline{\gamma}_{\hat{i}}}{(1-\mu_{i})\zeta_{ij}x}}_{y_{i}=0}
f_{Y_{i}}(y_{i})\nonumber\\
&\times F_{Z_{i}}\left(\frac{\overline{\gamma}_{\hat{i}}-(1-\mu_{i})\zeta_{ij}y_{i}x}{(1-\mu_{i})\zeta_{i\hat{j}}x}\right)dy_{i}\,dx.
\end{align}
Now, after applying some mathematical formulations and utilizing \cite[eq. 3.471.9]{math}, \eqref{kmklw} can be expressed as
\begin{align}\label{eije}
\!F_{\gamma_{i,\hat{i}}}(\overline{\gamma}_{\hat{i}})\!&=1-\sum^{m_{ij}-1}_{q=0}\frac{\left(\frac{m_{ij}\overline{\gamma}_{\hat{i}}}{\Omega_{ij}(1-\mu_{i})\zeta_{ij}}\right)^{q}}{\Gamma[m_{i\hat{i}}]q!}
\left(\frac{m_{i\hat{i}}}{\Omega_{i\hat{i}}}\right)^{m_{i\hat{i}}}\nonumber\\
&\times 2\left(\frac{m_{ij}\overline{\gamma}_{\hat{i}}\Omega_{i\hat{i}}}{\Omega_{ij}m_{i\hat{i}}(1-\mu_{i})\zeta_{ij}}\right)^{\frac{m_{i\hat{i}}-q}{2}}\nonumber\\
&\times\mathcal{K}_{m_{i\hat{i}}-q}\left(2\sqrt{\frac{m_{ij}\overline{\gamma}_{\hat{i}}m_{i\hat{i}}}{\Omega_{ij}(1-\mu_{i})\zeta_{ij}\Omega_{i\hat{i}}}}\right)
-\frac{\left(\frac{m_{ij}}{\Omega_{ij}}\right)^{m_{ij}}}{\Gamma[m_{ij}]}\nonumber\\
&\times\int^{\infty}_{x=0}f_{X}(x){\mathrm e}^{-\frac{m_{i\hat{j}}
		\overline{\gamma}_{\hat{i}}}{\Omega_{i\hat{j}}(1-\mu_{i})\zeta_{i\hat{j}}x}}
\sum^{m_{i\hat{j}}-1}_{k=0}
\frac{\left(\frac{m_{i\hat{j}}}{\Omega_{i\hat{j}}(1-\mu_{i})\zeta_{i\hat{j}}x}\right)^{k}}{k!}\nonumber\\
&\times\sum^{k}_{s=0}\binom{k}{s}\overline{\gamma}_{\hat{i}}^{s}(-1)^{k-s}((1-\mu_{i})\zeta_{ij}x)^{k-s}\nonumber\\
&\times\int^{\frac{\overline{\gamma}_{\hat{i}}}{(1-\mu_{i})\zeta_{ij}x}}_{y_{i}=0}y^{m_{ij}-1+k-s}_{i}{\mathrm e}^{\left(\frac{m_{i\hat{j}}\zeta_{ij}}
	{\Omega_{i\hat{j}}\zeta_{i\hat{j}}}-\frac{m_{ij}}{\Omega_{ij}}\right)y_{i}}dy_{i}dx.
\end{align}
On applying Maclaurin series expansion to the term ${\mathrm e}^{\frac{m_{i\hat{j}}\zeta_{ij}}
	{\Omega_{i\hat{j}}\zeta_{i\hat{j}}}}$ in \eqref{eije} and using  \cite[eqs. 3.471.9]{math}, the desired solution is obtained as given in Lemma \ref{lem3}.
	\end{IEEEproof}
\end{lemma}

On inserting \eqref{saljd} and \eqref{sioio} in \eqref{oopd}, one can obtain the desired OP expression for the IoT system.

\subsection{System Throughput}
For a delay-limited scenario, the system throughput for the considered cognitive radio network can be defined as the sum of average target rates of two primary and two IoT transmissions that can be successfully achieved over fading channels \cite{LiuIET}, \cite{Nasir2}. Therefore, one can express the system throughput as
\begin{align}\label{throughput}
\mathcal{S}_{\mathcal{T}}=\mathcal{S}_{\rm p}+\mathcal{S}_{\rm IoT}
\end{align}
where $\mathcal{S}_{\rm p}$ and $\mathcal{S}_{\rm IoT}$ are throughputs of the primary and IoT systems, respectively. By utilizing the expressions of individual OP of primary and IoT links, the expressions of $\mathcal{S}_{\rm p}$ and $\mathcal{S}_{\rm IoT}$ are given as \cite{Nasir2}
\begin{align}\label{throughputP}
\mathcal{S}_{\rm p}=\frac{(1-\beta)}{3}\big[(1-\mathcal{P}_{\textmd{out},a})r_{a}+(1-\mathcal{P}_{\textmd{out},b})r_{b}\big]
\end{align}
and
\begin{align}\label{throughputI}
\mathcal{S}_{\rm IoT}=\frac{(1-\beta)}{3}\big[(1-\mathcal{P}_{\textmd{out},1})r_{1}+(1-\mathcal{P}_{\textmd{out},2})r_{2}\big]
\end{align}
where $\mathcal{P}_{\textmd{out},a}$ $\&$ $\mathcal{P}_{\textmd{out},b}$ are given in \eqref{oopc} and $\mathcal{P}_{\textmd{out},1}$ $\&$ $\mathcal{P}_{\textmd{out},2}$ are defined in \eqref{oopd}.

\subsection{Energy Efficiency}
Designing energy efficient wireless networks is getting much attention nowadays and it proceeds towards relying on green communication technologies \cite{2Huang2018}. Consequently, it is relevant to examine the energy efficiency and investigate the impact of different parameters. The energy efficiency of the considered EH-based cognitive radio can be defined, based on the classical definition, as the ratio of total amount of data delivered to the total amount of consumed energy \cite{LiuIET}. For a delay-limited scenario, the expression of energy efficiency is
\begin{align}\label{sdnkad}
\eta^{\textmd{TS}}_{\textmd{EE}}=\frac{\mathcal{S}_{\mathcal{T}}}{\left(\frac{1+2\beta}{3}\right)(P_{a}+P_{b})}
\end{align}
where $\mathcal{S}_{\mathcal{T}}$ is given in \eqref{throughput}.

\subsection{Effective Spectrum Sharing}\label{spectrum}
When both PUs communicate to each other directly without receiving relay cooperation of IoDs, the achievable rate $\mathcal{R}^{\textmd{\tiny{D}}}_{\hat{j},j}$ at the primary nodes can be expressed as
\begin{align}\label{lkkr}
\mathcal{R}^{\textmd{\tiny{D}}}_{\hat{j},j}=\frac{1}{2}\log_{2}\left(1+\frac{ P_{\hat{j}}|h_{\hat{j},j}|^{2}}{\sigma^{2}}\right).
\end{align}
Here the pre-log factor $1/2$ appears due to the fact that two successive phases are required to realize end-to-end transmissions. Hereby, the OP for this direct transmission link can be given as
\begin{align}\label{iwhw}
\mathcal{P}^{\textmd{\tiny{D}}}_{\textmd{out},j}&=\textmd{Pr}\left[\frac{ P_{\hat{j}}|h_{\hat{j},j}|^{2}}{\sigma^{2}}<\widetilde{\gamma}_{j}\right]
=\frac{\Upsilon\left[m_{\hat{j}j},\frac{m_{\hat{j}j}\widetilde{\gamma}_{j}\sigma^{2}}{\Omega_{\hat{j}j} P_{\hat{j}}}\right]}{\Gamma[m_{\hat{j}j}]},
\end{align}
where $\widetilde{\gamma}_{j}=2^{2r_{j}}-1$. In fact, the spectrum sharing for IoT devices can be permissible until it does not affect the required outage performance of the primary system. However, for effective spectrum sharing, the primary links of the considered EH-enabled system should attain equal or lower OP than that of the direct transmissions (without spectrum sharing) of the primary system for the same predefined target rate \cite{Pandhari}, \cite{Pankaj}, i.e.,
\begin{align}\label{saeee}
\mathcal{P}_{\textmd{out},a}\leq\mathcal{P}^{\textmd{\tiny{D}}}_{\textmd{out},a}\,\,\textmd{and}\,\,\mathcal{P}_{\textmd{out},b}\leq\mathcal{P}^{\textmd{\tiny{D}}}_{\textmd{out},b}.
\end{align}
On utilizing \eqref{saeee}, one can obtain the critical value of power splitting factor (say $\mu^{\star}$) for which the system can offer effective spectrum sharing, where $\mu^{\star}\leq\mu_{i}$. Although the analytical evaluation of $\mu^{\star}$ directly from \eqref{saeee} appears mathematically intractable, numerical methods can be used to obtain the desired value.
\subsection{Average End-to-End Transmission Time}
The estimation of end-to-end transmission time for a packet to reach the destination is useful in the design of cognitive radio networks to meet latency requirements. According to Shannon's third theorem, the transmission time is inversely proportional to the achievable transmission rate of the corresponding channel \cite{FKhan}. Therefore, the time taken by a packet to arrive at the destination ${S}_{m}$ after transmitting from the source $S_{l}$ is given by
\begin{align}\label{adfee}
\Delta_{l,m}=\frac{\mathcal{L}}{\mathcal{W}\log_{2}(1+\gamma_{l,m})}=\frac{\widetilde{\mathcal{L}}}{\ln(1+\gamma_{l,m})}
\end{align}
where $\mathcal{L}$ is the length of the packet, $\mathcal{W}$ is the bandwidth of the channel,
and  $\widetilde{\mathcal{L}}=\mathcal{L}\ln(2)/\mathcal{W}$. Further, it is assumed that the transmission time and processing time of feedback/acknowledgment signals are negligible as compared to the packet transmission time and the transmitted packet arrives at the destination node before time-out \cite{NBMehta}, \cite{Sourabh}. With the considered system, the total transmission time is given as
\begin{align}\label{dookd}
\Delta_{j,\hat{j}}=\mathcal{T}_{\textmd{EH}}+\mathcal{T}_{\textmd{IP}}
\end{align}
where $\mathcal{T}_{\textmd{EH}}$ denotes the time taken for energy harvesting and $\mathcal{T}_{\textmd{IP}}$ represents the time taken for information processing and broadcasting. Since $\mathcal{T}_{\textmd{EH}}=\beta T$ and $\mathcal{T}_{\textmd{IP}}=(1-\beta) T$, the relationship between $\mathcal{T}_{\textmd{EH}}$ and $\mathcal{T}_{\textmd{IP}}$ is $\mathcal{T}_{\textmd{EH}}= \frac{\beta}{1-\beta}\mathcal{T}_{\textmd{IP}}$. Now, \eqref{dookd} can be expressed as
\begin{align}\label{dfwaaa}
\Delta_{j,\hat{j}}=\frac{1}{1-\beta}\mathcal{T}_{\textmd{IP}}.
\end{align}
For the considered relaying scheme, the average end-to-end transmission time from ${\sf PU}_{j}$ to ${\sf PU}_{\hat{j}}$ is determined as
{\small{\begin{align}\label{sdmdm}
\!\!\overline{\Delta}_{j,\hat{j}}&\!=\!\frac{1}{1\!-\!\beta}\Bigg(\textmd{Pr}[\mathcal{Q}_{i}]\left(1-\textmd{Pr}[\mathcal{Q}_{\hat{i}}]\right)\left(\mathds{E}\left[\max\left(\Delta_{j,i},\Delta_{j,\hat{i}}\right)\right]\!+\!\mathds{E}[\Delta_{i,\hat{j}}]\right)\nonumber\\
&+
\textmd{Pr}[\mathcal{Q}_{\hat{i}}]\left(1-\textmd{Pr}[\mathcal{Q}_{i}]\right)\left(\mathds{E}\left[\max\left(\Delta_{j,i},\Delta_{j,\hat{i}}\right)\right]+\mathds{E}[\Delta_{\hat{i},\hat{j}}]\right)\nonumber\\
&\!+\!\textmd{Pr}[\mathcal{Q}_{i}]\textmd{Pr}[\mathcal{Q}_{\hat{i}}]\!\!\left(\mathds{E}\left[\max\left(\Delta_{j,i},\Delta_{j,\hat{i}}\right)\right]\!+\!\mathds{E}[\Delta_{i,\hat{j}}]\!+\!\mathds{E}[\Delta_{\hat{i},\hat{j}}]\right)\!\!\Bigg)
\end{align}}}
where $i,\hat{i}\in\{1,2\}$ and $j,\hat{j}\in\{a,b\}$, with $i\neq\hat{i}$, $j\neq\hat{j}$. Furthermore, $\Delta_{j,i}$, $\Delta_{j,\hat{i}}$, $\Delta_{i,\hat{j}}$, and $\Delta_{i,j}$ can be obtained from \eqref{adfee} by inserting respective expressions of the instantaneous SNRs. If any relaying device fails to decode primary signals in the MAC phase, then it aborts broadcasting of the combined signal in the assigned BC phase. As a result, a limited feedback signal will be transmitted by that device to acknowledge all the corresponding nodes. For brevity, it is assumed that the time taken in this process is negligible \cite{Sourabh}. For the case when both relaying IoDs successfully decode the primary signals and broadcast in two consecutive BC phases, the average end-to-end transmission time will predominantly depend on the receiving time of both signal copies at the destination node for exploiting the selection combining technique. Note that, the exact derivation of \eqref{sdmdm} is highly intractable due to the involved complexity. As such, \eqref{sdmdm} is computed with Monte Carlo simulation in MATLAB. On the other hand, the  average end-to-end transmission time for the direct transmission from ${\sf PU}_{j}$ to ${\sf PU}_{\hat{j}}$ can be formulated as
\begin{align}\label{eemsl}
\overline{\Delta}^{\footnotesize{\textmd{(D)}}}_{j,\hat{j}}=\mathds{E}\left[\frac{\widetilde{\mathcal{L}}}{\ln\left(1+\gamma^{\footnotesize{\textmd{D}}}_{j,\hat{j}}\right)}\right]
\end{align}
For numerical results, \eqref{sdmdm} and \eqref{eemsl} are used to compare the end-to-end transmission time of our proposed scheme and the direct-link transmission time.

\begin{table}[t]
	\centering
	\caption{Number of terms required in infinite series of \eqref{oopc} and \eqref{oopd} for achieving  accuracy
		up to first seven decimal places.}
	\label{table2}
	\begin{tabular}{c|c|c|c|c}
		\hline\hline
		\multirow{2}{*}{Index $p$} & \multicolumn{2}{l|}{$\,\,\,\quad\quad\,\,\mathcal{P}_{\textmd{out},a}$ in \eqref{oopc}} & \multicolumn{2}{l}{$\quad\quad\mathcal{P}_{\textmd{out},1}$ in \eqref{oopd}} \\ \cline{2-5}
		& ${\rm SNR}=5$dB         & $10$dB       & $10$dB       & $15$dB        \\ \hline
		1                      & 0.720438        & 0.129212       & 0.408842       & 0.0859022       \\ \hline
		2                      & 0.618687        & 0.108401       & 0.359605       & 0.0674041       \\ \hline
		3                      & 0.644159        & 0.111233       & 0.338323       & 0.0610553       \\ \hline
		4                      & 0.639092        & 0.110922       & 0.329713       & 0.0588903       \\ \hline
		5                      & 0.639904        & 0.11095        & 0.326344       & 0.0581447       \\ \hline
		6                      & 0.639794        & 0.110947       & 0.325048       & 0.0578843       \\ \hline
		7                      & 0.639807        & 0.110948       & 0.324553       & 0.057792        \\ \hline
		8                      & 0.639806        & 0.110948       & 0.324365       & 0.0577589       \\ \hline
		9                      & 0.639806        & 0.110948       & 0.324293       & 0.0577469       \\ \hline
		10                     & 0.639806        & 0.110948       & 0.324266       & 0.0577425       \\ \hline
		11                     & 0.639806        & 0.110948       & 0.324255       & 0.0577408       \\ \hline
		12                     & 0.639806        & 0.110948       & 0.324251       & 0.0577402       \\ \hline
		13                     & 0.639806        & 0.110948       & 0.324250       & 0.0577400       \\ \hline
		14                     & 0.639806        & 0.110948       & 0.324249       & 0.0577399       \\ \hline
		15                     & 0.639806        & 0.110948       & 0.324249       & 0.0577399       \\ \hline \hline
	\end{tabular}
\end{table}

\section{Numerical and Simulation Results}\label{numerical}
\begin{figure}[t!]
	\centering
	\includegraphics[width=3.6in]{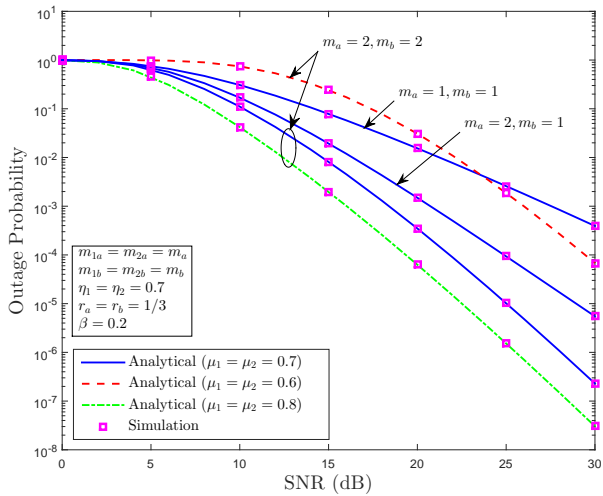}
	\caption{OP versus SNR curves for ${\sf PU}_{b}\rightarrow {\sf PU}_{a}$ link.}
	\label{fig3}
\end{figure}
\begin{figure}[t!]
	\centering
	\includegraphics[width=3.6in]{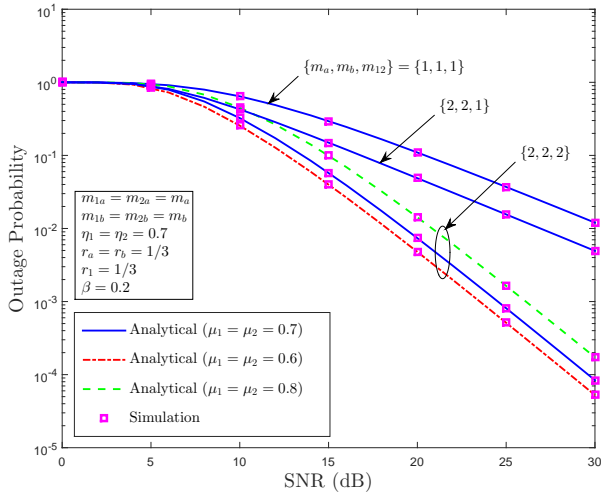}
	\caption{OP versus SNR curves for ${\sf IoD}_{2}\rightarrow {\sf IoD}_{1}$ link.}
	\label{fig4}
\end{figure}
\begin{figure}[t!]
	\centering
	\includegraphics[width=3.6in]{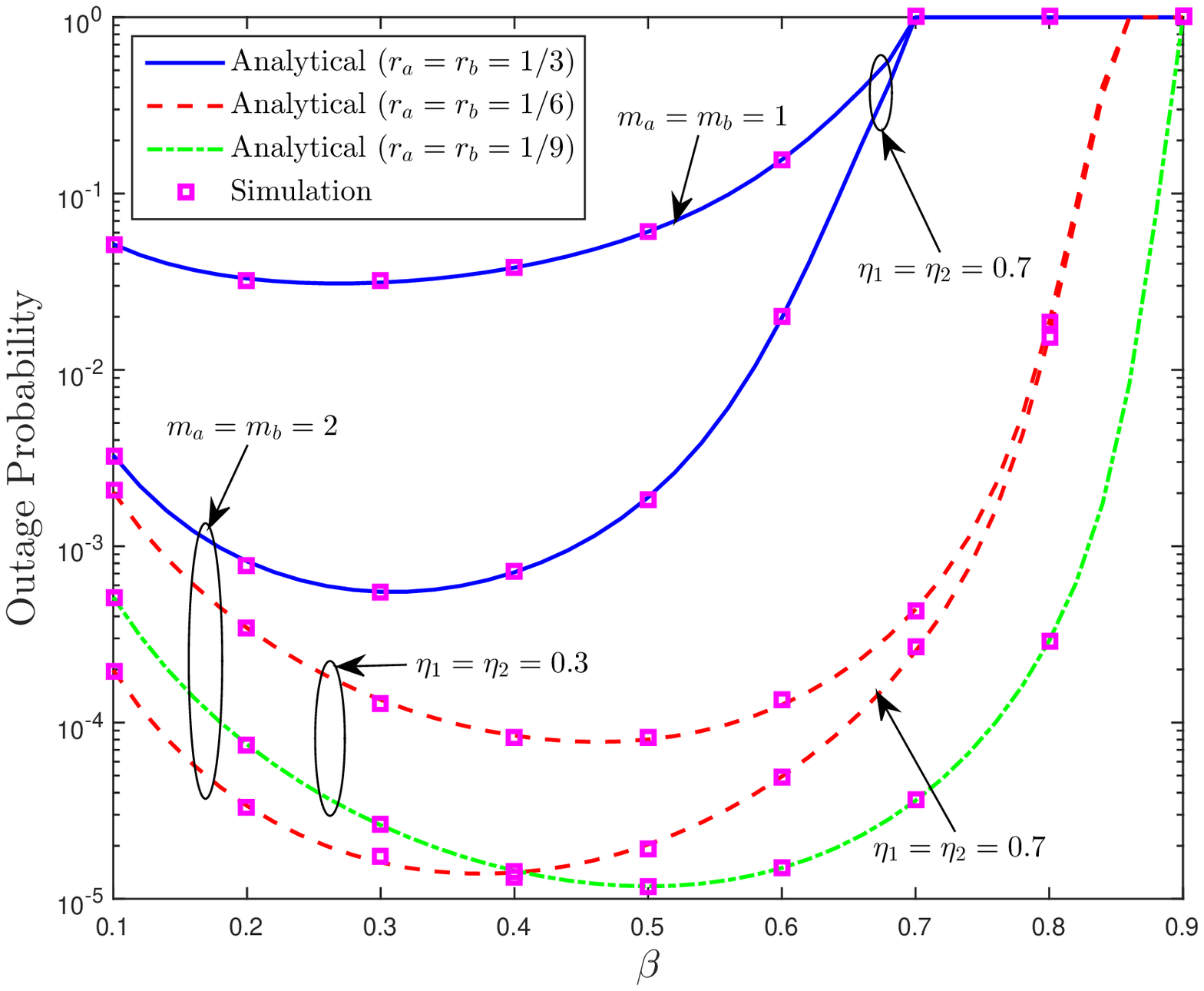}
	\caption{OP versus $\beta$ curves for ${\sf PU}_{b}\rightarrow {\sf PU}_{a}$ link.}
	\label{fig5}
\end{figure}
\begin{figure}[t!]
	\centering
	\includegraphics[width=3.6in]{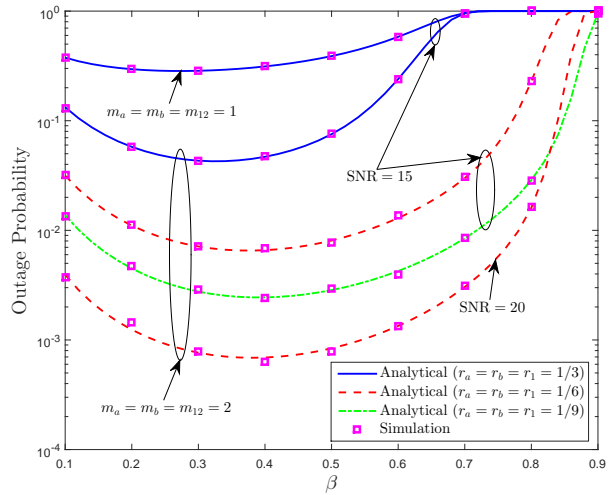}
	\caption{OP versus $\beta$ curves for ${\sf IoD}_{2}\rightarrow {\sf IoD}_{1}$ link.}
	\label{fig6}
\end{figure}

\begin{figure}[t!]
	\centering
	\includegraphics[width=3.6in]{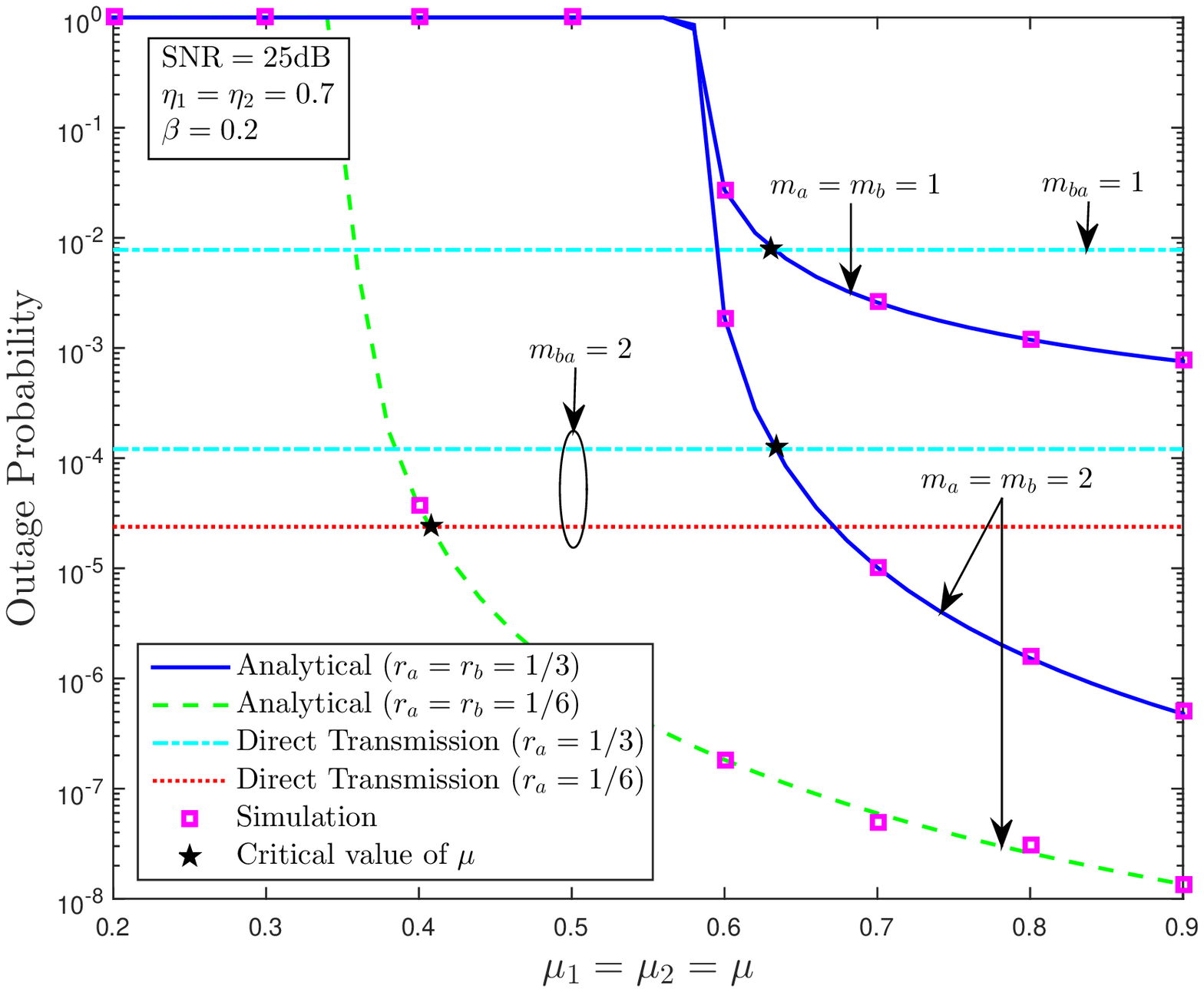}
	\caption{OP versus $\mu$ curves for ${\sf PU}_{b}\rightarrow {\sf PU}_{a}$ link.}
	\label{fig7}
\end{figure}

\begin{figure}[t!]
	\centering
	\includegraphics[width=3.6in]{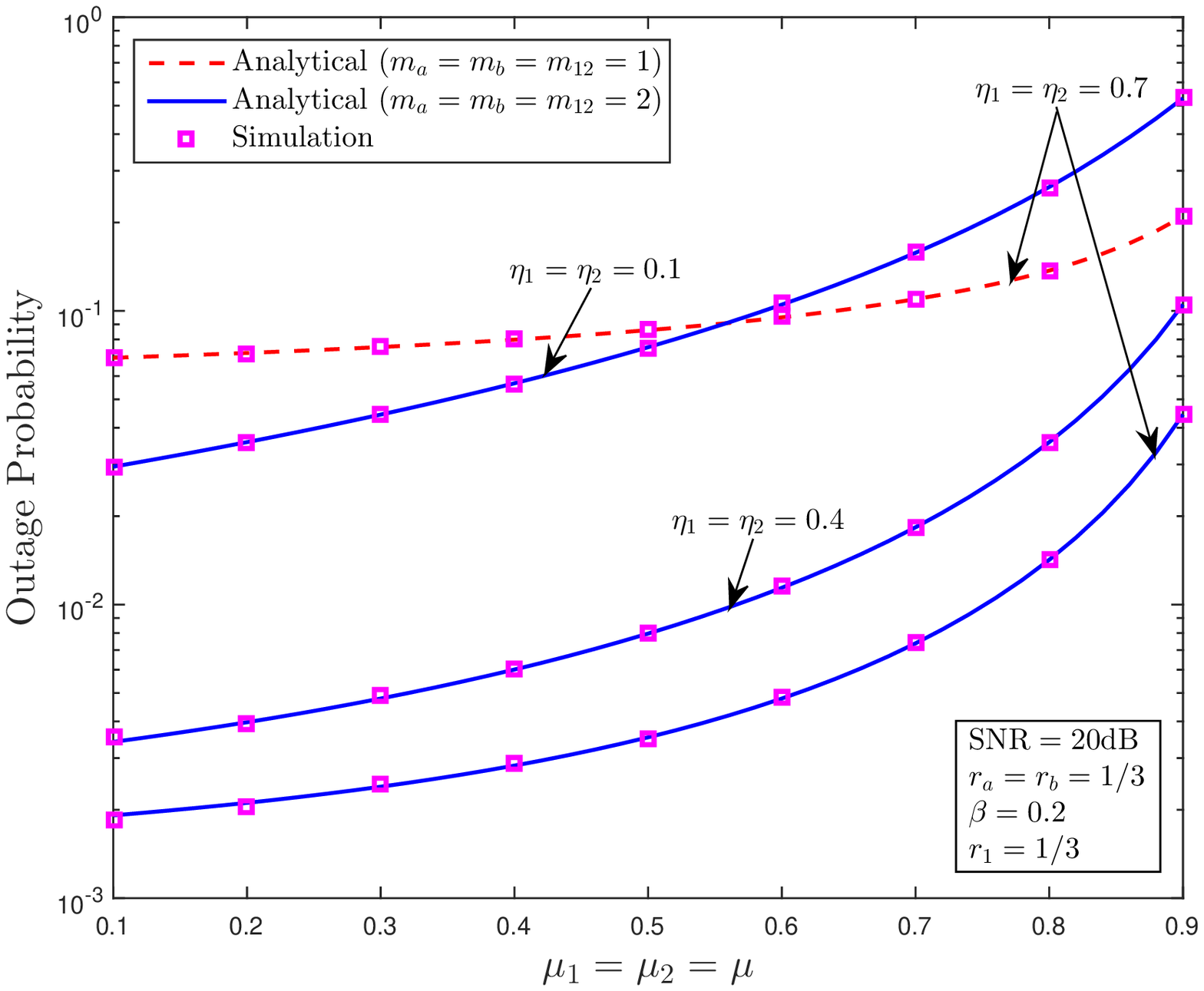}
	\caption{OP versus $\mu$ curves for ${\sf IoD}_{2}\rightarrow {\sf IoD}_{1}$ link.}
	\label{fig8}
\end{figure}

\begin{figure}[t!]
	\centering
	\includegraphics[width=3.6in]{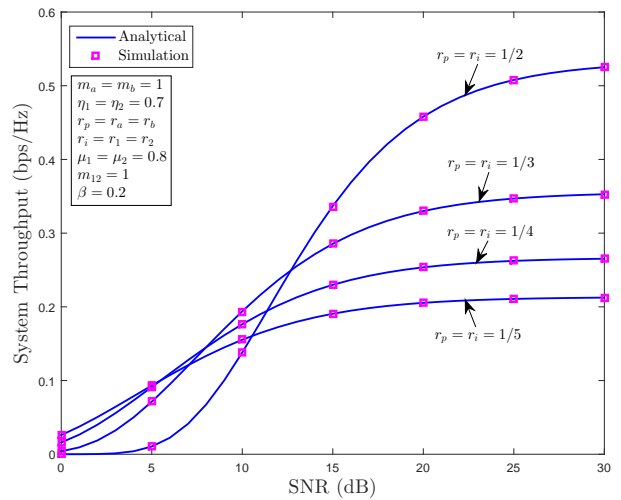}
	\caption{System throughput versus SNR curves with different target rates.}
	\label{fig9}
\end{figure}

\begin{figure}[t!]
	\centering
	\includegraphics[width=3.6in]{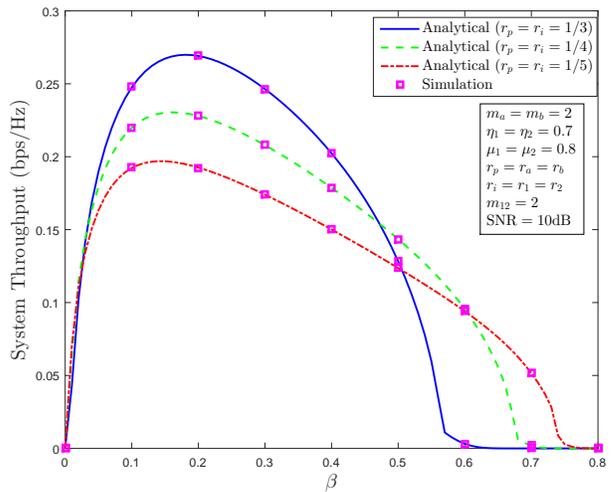}
	\caption{System throughput versus $\beta$ curves with different parameters.}
	\label{fig10}
\end{figure}

\begin{figure}[t!]
	\centering
	\includegraphics[width=3.6in]{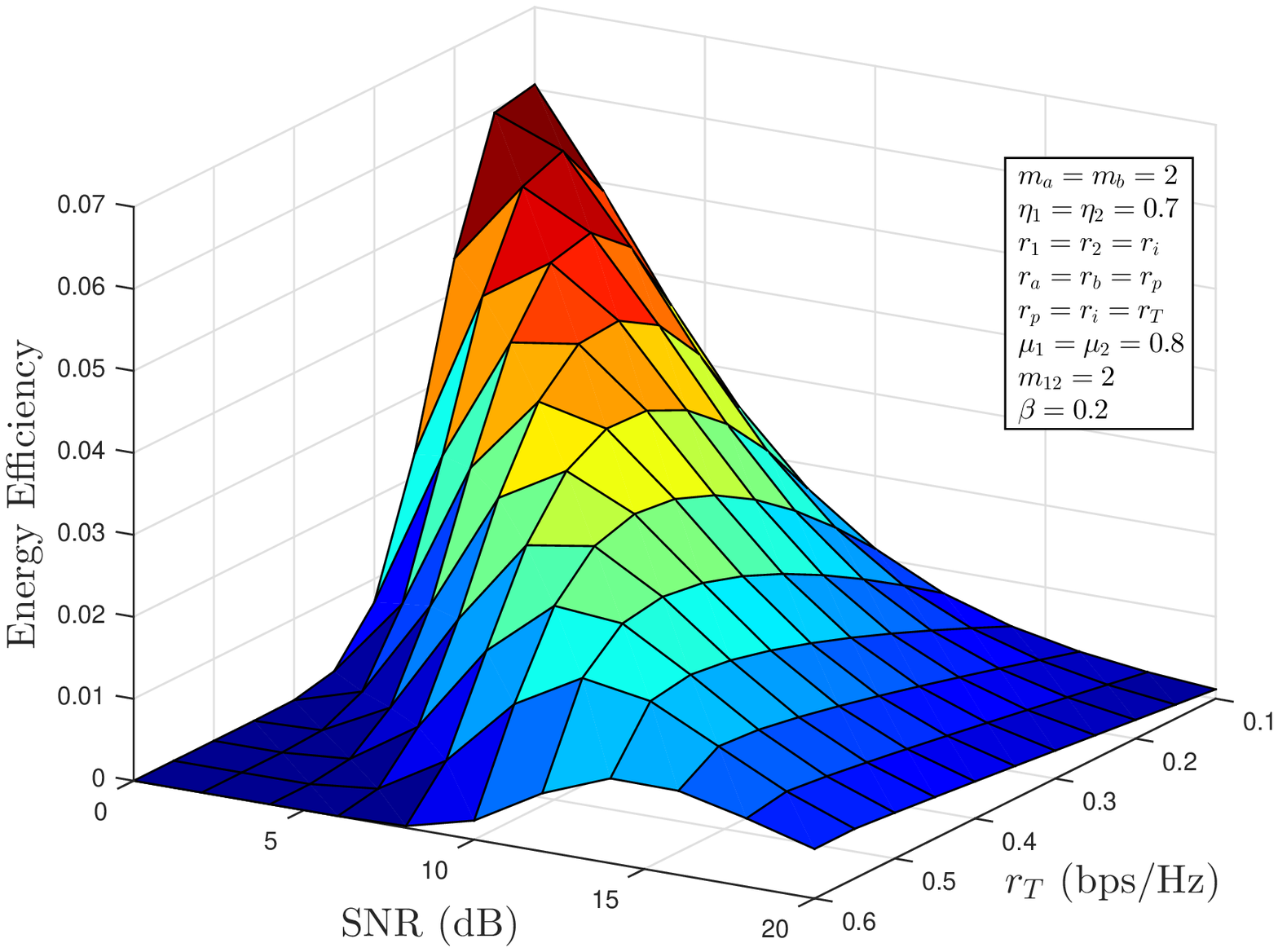}
	\caption{Energy efficiency versus SNR and  target rate.}
	\label{fig11}
\end{figure}

This section presents numerical and simulation results and discusses the impact of key system/channel parameters on the performance of the considered SWIPT-enabled cognitive radio system. For all the numerical results, it is assumed that $P_{a}=P_{b}=P$ and define ${P}/{\sigma^{2}}$ as the transmit SNR. Further, this paper adopts a path-loss model where the variances of channel gains are defined in terms of the corresponding distances between two nodes and the path-loss exponent. As such, for links ${\sf PU}_{a}\rightarrow {\sf IoD}_{1}$ and ${\sf PU}_{a}\rightarrow {\sf IoD}_{2}$, the variances of channel coefficients are defined as $\Omega_{1a}=d_{1a}^{-\nu}$ and $\Omega_{2a}=d_{2a}^{-\nu}$, respectively. Similarly, for ${\sf PU}_{b}\rightarrow {\sf IoD}_{1}$, ${\sf PU}_{b}\rightarrow {\sf IoD}_{2}$, and ${\sf IoD}_{2}\rightarrow {\sf IoD}_{1}$ links, the variances of channel coefficients are $\Omega_{1b}=d_{1b}^{-\nu}$, $\Omega_{2b}=d_{2b}^{-\nu}$, and $\Omega_{12}=d_{12}^{-\nu}$, respectively. All simulation results were obtained by considering that $d_{1a}=d_{2a}=1$, $d_{1b}=d_{2b}=0.9$, and $d_{12}=1$ with path-loss exponent $\nu=3$. Moreover, fading severity parameters are set as $m_{1a}=m_{2a}=m_{a}$ and $m_{1b}=m_{2b}=m_{b}$. The values of other system/channel parameters vary in different figures and are specified therein. For obtaining numerical values in Table \ref{table2}, the parameters are set as $m_{a}=m_{b}=m_{12}=2$, $\beta=0.2$, $\eta_{1}=\eta_{2}=0.7$, and $r_{a}=r_{b}=r_{1}=r_{2}=1/3$. As depicted in Table \ref{table2}, the infinite series involved in \eqref{slkksdl} and \eqref{sioio} are truncated to include the first fifteen terms for achieving the sufficient accuracy (first seven decimal places) in all the analytical results.

\subsection{Outage Probability with SNR}
In obtaining Fig. \ref{fig3}, the system parameters are set as $\eta_{1}=\eta_{2}=0.7$, $\beta=0.2$, and $r_{a}=r_{b}=r_{\textmd{th}}=1/3$ bps/Hz. This figure shows the OP versus SNR curves for the primary link ${\sf PU}_{b}\rightarrow {\sf PU}_{a}$ of the considered system with various fading scenarios. All the analytical curves are in good consonance with the simulation results, which confirms the accuracy of the derived analytical expressions. It can be seen From Fig. \ref{fig3} that when the value of $m_{a}$ and/or $m_{b}$ increases from $1$ to $2$, the user OP performance of the primary system improves. From this, one can infer that the system experiences better OP performance with comparatively less severe fading conditions. Given that the OP performance improves with higher SNR values, a required SNR value can be identified to achieve a desired link reliability. Moreover, Fig. \ref{fig3} also shows OP curves for different values of power splitting factors $\mu_{1}$ and $\mu_{2}$. It can be observed that as the value of $\mu_{i}$ increases, the OP of the primary link also improves. This behavior is in agreement with the modeling of spectrum sharing system, where a higher value of $\mu_{i}$ represents that more power is assigned for primary transmissions.

Fig. \ref{fig4} plots the OP curves versus SNR of the IoT link ${\sf IoD}_{2}\rightarrow {\sf IoD}_{1}$ for different fading scenarios. For this figure, the target rate is set as $r_{1}=1/3$ and all other parameters are the same as in Fig. \ref{fig3}. As can be seen, all the simulation points are in perfect match with the corresponding analytical curves. Similar to Fig. \ref{fig3}, as the value of fading parameters increases, the OP of the IoT link also improves. On the other hand, when the value of $\mu_{i}$ increases, the corresponding OP of the IoT link degrades. This is because the power allocated for IoT transmissions is scaled by ($1-\mu_{i}$) term.

\subsection{Outage Probability with TS Factor}
For numerical investigation in Fig. \ref{fig5}, the parameter $\mu_{1}=\mu_{2}=0.9$, ${\rm SNR}=15$dB. This figure plots OP curves versus the TS factor $\beta$ for various fading scenarios and different values of energy conversion efficiency at the IoDs. From Fig. \ref{fig5}, one can see that for a given set of parameters, the primary system achieves the lowest OP at a certain value of $\beta$. If the value of $\beta$ increases or decreases from that value, the system OP performance degrades. For lower values of $\beta$, OP increases because less time is allocated for EH at IoDs and hence less transmit power available at IoDs. On the other hand, when the value of $\beta$ increases after a certain value, the OP also increases due to a drastic rise in the target SNR with the factor $2^{\frac{3r_{\textmd{th}}}{1-\beta}}-1$. Therefore, it is crucial to set an appropriate value of $\beta$ to get optimal OP performance. Moreover, when the target rate at PUs increases, the OP performance of the primary link degrades. This behavior shows the trade-off between link reliability and achievable data rate for the primary system. Energy conversion efficiency is another key factor in determining the OP performance of the primary link. Lower values of $\eta_{1}$ and $\eta_{2}$ lead to lower OP performance.

Fig. \ref{fig6} shows OP versus $\beta$ curves of the IoT link for various fading scenarios and SNR values. In this figure, the value of energy conversion efficiency is fixed as $\eta_{1}=\eta_{2}=0.7$ and power splitting factor is set as $\mu_{1}=\mu_{2}=0.7$. From this figure, one can see that as the target rate increases from $1/9$ to $1/3$, the OP performance of the IoT link degrades. However, this degradation in OP performance can be recovered if the value of SNR increases from $15$dB to $20$dB.

\subsection{Outage Probability with Spectrum Sharing Factor}
Section \ref{spectrum} highlights that for effective spectrum sharing, the value of power splitting factor $\mu_{i}$ should be chosen carefully. In addition, Figs. \ref{fig3} and \ref{fig4} also demonstrate that the value of power splitting factor has crucial impacts on the performance of both primary and IoT systems. Hereby, OP versus $\mu$ curves are plotted in Figs. \ref{fig7} and \ref{fig8} for the primary and IoT systems, respectively,  to show the feasible range and offer some insightful observations. In Fig. \ref{fig7}, the system parameters are set as $\beta=0.2$, $\eta_{1}=\eta_{2}=0.7$, and ${\rm SNR}=25$dB. From this figure, one can observe that for the effective spectrum sharing, the value of $\mu_{i}$ should be greater than a certain value. For determining that critical value of $\mu_{i}$, the solution of \eqref{saeee} is obtained using a numerical method. In Fig. \ref{fig7}, the critical value of $\mu_{i}$ ($\mu^{\star}$) can be referred as a point at which the OP of the primary link with the proposed scheme shows the same OP of direct transmission (shown by horizontal lines) curves. As such, for $\mu^{\star}<\mu_{i}$, the primary system exhibits better outage performance than that of the direct transmission. Consequently, the effective range for spectrum sharing can be given as $\mu^{\star}<\mu_{i}<1$. On the other hand, with the setting of $r_{\textmd{th}}=1/3$ bps/Hz and $\beta=0.2$, the feasible range of power splitting factor will be $0.58<\mu_{i}<1$ for enabling spectrum sharing. Below this value, the OP of the primary link becomes unity  as also highlighted by Lemma \ref{lem2}. From here, one can note that the minimum possible value of $\mu_{i}$ depends only on the TS parameter $\beta$ and target rates of the corresponding primary links. Further, in Fig. \ref{fig8}, the system parameters are set as  $\beta=0.2$, $r_{\textmd{th}}=1/3$, and ${\rm SNR}=20$dB. It can be seen from this figure that, as the value of $\mu_{i}$ increases, the OP performance of the IoT link degrades. Different from the primary link, the IoT link shows considerable OP performance for the entire range of $\mu_{i}$.

\begin{figure}[t]
\centering
\includegraphics[width=3.5in]{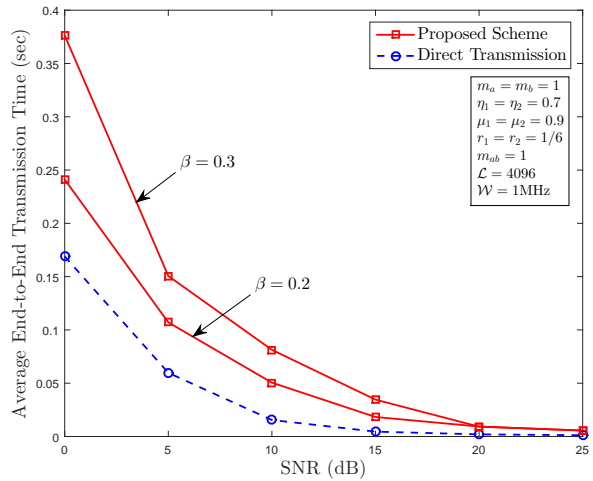}
\caption{Average end-to-end transmission time versus SNR curves for ${\sf PU_{a}}$ to ${\sf PU_{b}}$ link.}
\label{fig12}
\end{figure}

\subsection{System Throughput}
Fig. \ref{fig9} plots the system throughput versus SNR curves. Herein, the parameters are set as $m_{a}=m_{b}=m_{12}=1$, $\beta=0.2$, $\mu_{1}=\mu_{2}=0.8$ and $\eta_{1}=\eta_{2}=0.7$. Observe that, for low SNR values, the curves corresponding to higher target rates show lower system throughput as compared to the curves corresponding to lower target rates. This is due to the fact that in the low SNR region, as the value of target rate increases, the corresponding target SNR also increases, which degrades the OP performance of both systems. When the OP of both systems become higher, the system throughput decreases. On the contrary, in medium to high SNR region, the impact of degradation in OP performance is less as compared to the enhancement due to higher target rates. After a particular SNR value, the system throughput curves attain a saturation point that can be referred to as the maximum achievable throughput for the considered set of parameters.

The setting of TS factor is also crucial for system throughput performance. Fig. \ref{fig10} shows the system throughput versus $\beta$ curves. Herein, the parameters are set as $m_{a}=m_{b}=m_{12}=2$, $\eta_{1}=\eta_{2}=0.7$, $\mu_{1}=\mu_{2}=0.8$, and ${\rm SNR}=10$dB.  As expected, the curves corresponding to higher target rates attain the maximum achievable throughput in the range   $0<\beta<0.5$. For the case when target rate is $r_{\textmd{th}}=1/3$, the system achieves the maximum throughput at $\beta=0.18$ for the considered set of parameters. When the target rates decrease, the value of $\beta$ at which the system attains the maximum throughput also shifts towards lower values.

\subsection{Energy Efficiency}
To reveal the impact of different parameters on the overall energy efficiency of the considered system, Fig. \ref{fig11} plots the energy efficiency versus SNR and target rate. Here, the parameters are set as $\beta=0.2$, $\eta_{1}=\eta_{2}=0.7$, $\mu_{1}=\mu_{2}=0.8$, and the target rates of both primary and IoT systems are assumed to be equal. From this figure, one can see that with a lower target rate, the system achieves significant energy efficiency at lower SNR values. For example, when the target rate is  0.1 bps/Hz, the maximum energy efficiency is achieved at $0$dB. On the other hand, when the target rate is higher, the system attains better energy efficiency from medium to high SNR regime. Based on this observation, one can infer that the maximum energy efficiency can be attained at specific values of SNRs only, and that depends on the required target rates.  As such, when the target rate increases from a lower value, the SNR value for which the system achieves the maximum energy efficiency also shifts towards the higher value.

\subsection{Average Transmission Time}
Fig. \ref{fig12} plots the average end-to-end transmission times versus the transmit SNR for the proposed relaying scheme and direct transmission (without spectrum sharing). For the results in this figure, the parameters are set as $\mathcal{L}=4096$, $\mathcal{W}=1$ MHz, $\eta_{1}=\eta_{2}=0.7$, $\mu_{1}=\mu_{2}=0.9$, $r_{1}=r_{2}=1/6$ bps/Hz, and $m_{a}=m_{b}=m_{ab}=1$. As naturally expected, the end-to-end transmission time of the proposed scheme is higher than the transmission time of direct transmission. However, the absolute transmission times and their difference quickly decrease as the SNR increases. This small drawback should be easily outweighted by the superiority of the proposed scheme with regard to other important performance metrics, including spectral efficiency, link reliability, and energy efficiency.

\section{Conclusion}\label{conclusion}
This paper proposed a SWIPT-based spectrum sharing scheme to enable IoT communications in the licensed spectrum and to realize the primary communications with improved link reliability. A pair of SWIPT-based IoDs has been considered for providing relay assistance to primary transmission by applying decode-and-forward operation. First, this paper analyzed the outage performance of both primary and IoT systems with the proposed scheme under Nakagami-$m$ fading. Then, it formulated the expressions of energy efficiency and system throughput. Further, it discussed the condition for spectrum sharing for which the OP performance of the proposed scheme is equal or lower than that of the direct transmission. Numerical and simulation results elucidated the accuracy of all the derived expressions and highlighted the impacts of some critical design parameters, e.g., power splitting factor and time switching factor, on the system performance. Above all, this work incorporated the concept of the cognitive radio system, SWIPT, and spectral efficient relaying for the deployment of future IoT systems.

\end{document}